\documentclass[aps,prapplied,twocolumn,superscriptaddress,
floatfix,longbibliography]{revtex4-2}

\usepackage{graphicx}
\usepackage{amsmath,amssymb}
\usepackage{bm}
\usepackage{physics}
\usepackage{array}
\usepackage[table,dvipsnames]{xcolor}
\usepackage{tikz}
\usetikzlibrary{shapes.geometric,arrows.meta,positioning,shadows,calc,plotmarks}
\usepackage[colorlinks=true,linkcolor=blue,citecolor=blue,urlcolor=blue]{hyperref}
\usepackage{caption}
\usepackage{subcaption}
\usepackage{pgfplots}
\pgfplotsset{compat=1.18}
\usepgfplotslibrary{colormaps,groupplots}
\usepackage{colortbl}
\usepackage{booktabs}
\usepackage{multirow}
\usepackage{tabularx}
\usepackage{mdframed}

\definecolor{bottleamber}{HTML}{FFF3CD}
\definecolor{lowcolor}{RGB}{255,200,200}
\definecolor{highcolor}{RGB}{200,255,200}
\definecolor{qrcBlue}{HTML}{2166AC}
\definecolor{qrcGreen}{HTML}{4DAC26}
\definecolor{qrcRed}{HTML}{D6604D}
\definecolor{heatLo}{HTML}{D73027}
\definecolor{heatHi}{HTML}{1A9850}
\definecolor{cmHi}{HTML}{2166AC}
\definecolor{chanceGrey}{HTML}{777777}
\definecolor{QPurple}{RGB}{88,28,135}
\definecolor{QViolet}{RGB}{124,58,237}
\definecolor{QBlue}{RGB}{30,64,175}
\definecolor{QCyan}{RGB}{6,182,212}
\definecolor{QTeal}{RGB}{20,184,166}
\definecolor{QGreen}{RGB}{22,163,74}
\definecolor{QGold}{RGB}{217,119,6}
\definecolor{QOrange}{RGB}{234,88,12}
\definecolor{QRed}{RGB}{185,28,28}
\definecolor{QDark}{RGB}{15,23,42}
\definecolor{QLight}{RGB}{226,232,240}
\definecolor{QAccentA}{RGB}{56,189,248}
\definecolor{HF0}{RGB}{215,48,39}
\definecolor{HF1}{RGB}{244,109,67}
\definecolor{HF2}{RGB}{254,224,139}
\definecolor{HF3}{RGB}{116,196,118}
\definecolor{HF4}{RGB}{35,139,69}

\pgfplotsset{
  qrc/.style={
    width=\linewidth, font=\small,
    grid=major, grid style={dotted,gray!35},
    axis line style={QDark}, tick style={color=QDark},
    label style={font=\small\bfseries,color=QDark},
    title style={font=\small\bfseries,color=QDark},
    legend style={draw=QLight,fill=white,font=\scriptsize,rounded corners=2pt},
  }
}

\newcommand{\fclr}[1]{%
  \ifdim #1 pt < 0.40 pt \cellcolor{HF0!50}%
  \else\ifdim #1 pt < 0.55 pt \cellcolor{HF1!50}%
  \else\ifdim #1 pt < 0.70 pt \cellcolor{HF2!80}%
  \else\ifdim #1 pt < 0.80 pt \cellcolor{HF3!60}%
  \else \cellcolor{HF4!55}%
\fi\fi\fi\fi#1}

\newcommand{\similaritycell}[1]{%
  \pgfmathparse{int(round(#1*100))}%
  \edef\simgreen{\pgfmathresult}%
  \edef\simcolspec{green!\simgreen!red!30!white}%
  \expandafter\cellcolor\expandafter{\simcolspec}%
  \pgfmathprintnumber[fixed,precision=6]{#1}%
}

\newcommand{\nn}{\hat{n}}
\newcommand{\Ctilde}{\tilde{C}}
\newcommand{\kvec}{\bm{k}}
\newcommand{\rvec}{\bm{r}}

\begin{document}

\title{Rydberg Vision via frugal Quantum Image Fingerprinting}

\author{Vikrant Sharma}
\email{vikrant@dei.ac.in}
\affiliation{Electrical Engineering Department,
Dayalbagh Educational Institute, Agra, India}

\author{Neel Kanth Kundu}
\email{neelkanth@iitd.ac.in}
\affiliation{Centre for Applied Research in Electronics (CARE),
Indian Institute of Technology Delhi, New Delhi, India}

\date{\today}

\begin{abstract}
  Gate-based quantum image processing is constrained by qubit scarcity
  and the high overhead of quantum state preparation, limiting its
  applicability to realistic geometric data. We introduce a
  quantum-native framework for image matching on neutral-atom analog
  quantum computers that advances our earlier Sparse-Dots Representation
  (SDR) approach. A classical pre-processing pipeline---Sobel edge
  extraction followed by the Ramer--Douglas--Peucker (RDP)
  algorithm---converts an input image into a geometrically faithful
  Sparse-Dots point cloud of substantially fewer atoms. This atom layout
  is virtually embedded into the programmable tweezer array of QuEra's
  Aquila device via its Bloqade SDK, where the image geometry is encoded
  physically in the distance-dependent van der Waals interaction term of
  the Rydberg Hamiltonian. After time-evolution, we extract the many-body
  fingerprint of each image using two observables---the
  Pearson-normalized two-site correlation matrix which encodes the
  blockade-induced correlation structure of the quantum state, and the
  two-dimensional static structure factor evaluated on a fixed wavevector
  grid, yielding a fingerprint vector of constant length regardless of
  atom count. In Stage~1, image matching is performed by cosine
  similarity on the fingerprint vectors, a scale-invariant metric
  appropriate for Fourier-domain descriptors. In Stage~2, this approach
  is extended to quantum reservoir computing~(QRC) to enable machine
  learning via dramatically reduced training data and training cycles, as
  a preliminary proof-of-concept. Simulations using the Bloqade software
  stack confirm successful matching of industrial objects, often with
  fewer than 24 atoms. To our knowledge, this constitutes the first
  application of the static structure factor---a condensed-matter quantum
  observable---as an image retrieval descriptor in an analog quantum
  computing context.
\end{abstract}

\maketitle

\section{Introduction}
\label{sec:intro}

Human visual recognition is remarkably energy and data efficient.
In familiar environments, individuals are often identified without
reliance on full-frontal facial information, instead using coarse
geometric cues such as body build, posture, or partial views.
This process is robust to occlusion and illustrates that accurate
recognition does not require complete or high-resolution visual data.

Motivated by this observation, we explore whether analogous principles
can be exploited for quantum-native processing of geometric data on
current quantum hardware.
Rather than encoding full pixel-level information, we focus on sparse
geometric representations that preserve structural content while
discarding redundant detail.
Such representations naturally reduce quantum resource requirements,
making them well suited for the noisy intermediate-scale quantum (NISQ)
regime.
By contrast, existing approaches to quantum image processing
predominantly rely on digital encodings of pixel data into quantum
states, which incur substantial state-preparation overhead and scale
poorly under current qubit constraints~\cite{su2020}.

In this work, we introduce a quantum-native framework for encoding and
processing sparse geometric data directly into interacting many-body
quantum systems.
Instead of mapping images onto digital basis states, geometric
information is embedded through the sparse spatial arrangement of
neutral atoms, allowing native interatomic interactions to encode
structural relationships.
Under global Rydberg driving, these atomic configurations evolve into
correlated quantum states whose connected density correlations reflect
the underlying geometry.
We show that these emergent many-body correlations form compact and
experimentally accessible signatures of sparse visual structures.
By comparing correlation signatures across different inputs, robust
matching can be achieved using fixed Hamiltonian evolution and
projective measurements alone, without explicit digital quantum circuits
or classical feature extraction.
\textbf{The Hamiltonian itself is the Interferometer}.
These results establish many-body correlations as a natural similarity
metric for geometric data on analog quantum devices and highlight the
potential of neutral-atom platforms as resource-efficient,
quantum-native alternatives to classical image processing pipelines.

\subsection{Quantum Image Processing}

Quantum image processing (QImP)~\cite{yao2017} exploits quantum
mechanical properties---superposition, entanglement, and interference---
to represent and manipulate images.
Through massive parallel state evolution, QImP can provide advantages
over classical image processing for selected operations.
Quantum image representation (QImR) is the foundational step: it
determines which processing tasks are feasible and at what cost, as it
governs both the nature of image encoding and the realizable processing
tasks.
Edge extraction is a fundamental operation within QImP.
With the rapid growth of high-resolution and high-throughput image data,
large-scale real-time classical edge processing increasingly faces
challenges in energy consumption, latency, and system-level
scalability---thereby motivating quantum and hybrid processing
strategies.
Analog quantum platforms offer an alternative paradigm in which
physical interactions and interference directly generate expressive
similarity measures, without explicit feature construction or
optimization loops.
Emergent many-body correlations thus provide a natural substrate for
quantum-native pattern recognition, and suggest a pathway toward
reservoir-style learning on near-term neutral-atom hardware.

\subsection{Why Quantum Machine Learning for Image Processing?}

Machine learning (ML) encompasses computational methods for extracting
patterns from data, including classification, regression, clustering,
and related tasks.
Classical ML methods, particularly kernel-based classifiers, often face
fundamental scalability limitations when applied to data with highly
nonlinear and high-dimensional structure, as the explicit construction
and evaluation of complex feature spaces rapidly becomes computationally
prohibitive.
Quantum ML (QML)~\cite{schuld2019,havlicek2019,liu2021} offers a principled
alternative by exploiting the exponentially large Hilbert space of a
quantum system as a natural feature space, enabling efficient
realization of rich nonlinear kernel functions via quantum state
preparation and entangling operations~\cite{liu2021}.
As demonstrated by Havl\'{i}\v{c}ek et al.~\cite{havlicek2019},
quantum-enhanced feature maps can induce class separability that is
difficult or infeasible to replicate using classical kernels of
comparable computational cost.
Certain subroutines in AI/ML---linear algebra, kernel methods,
sampling, and combinatorial optimisation---admit quantum algorithms with
provable speedups, motivating the study of quantum-native learning
models that leverage interference and entanglement to construct
expressive similarity measures.
Early evidence suggests that quantum annealing and analog quantum computers can
be integrated into quantum-classical AI workflows, potentially enhancing
model efficiency and reducing energy
consumption~\cite{kornjaca2024,fujii2017}.

\paragraph{Energy efficiency.}
QuEra's \textit{Aquila}~\cite{wurtz2023}, a neutral-atom quantum device
with up to 256 programmable atom qubits, typically operates on less
than 7\,kW for the full system---according to industry estimates, less
than 0.05\% of the power required by a classical AI
supercomputer~\cite{boger2023}.
While control electronics, vacuum systems, and laser-stabilization
infrastructure contribute a broadly fixed overhead that does not vanish
at zero atoms, the qubit-related power costs (laser drive and atom
trapping) do not scale strongly with atom count in neutral-atom
architectures.
This is in marked contrast to classical computing, where active
transistor count drives energy consumption near-linearly with the
number of processing elements, making neutral-atom devices particularly
attractive for qubit-count scale-up.
For comparison, gate-based quantum computers from IBM, Rigetti and Google
consume approximately 10--25\,kW~\cite{boger2023}.

\paragraph{Native combinatorial optimization.}
The device provides hardware-native heuristic solutions to Maximum
Independent Set (MIS) problems using the Rydberg
blockade~\cite{ebadi2022}---a task for which even heuristic classical
solvers (simulated annealing, greedy algorithms) can require hours to
days on graphs with hundreds of nodes.
MIS is directly relevant to ML tasks including feature selection,
graph-based clustering, and resource allocation in networks.

\paragraph{Quantum reservoir computing.}
Quantum reservoir computing (QRC) on such hardware requires as few as
$\sim$300 training samples, avoids gradient descent entirely, and
therefore does not suffer from the vanishing gradient
problem~\cite{kornjaca2024,beaulieu2024}.
Once an image is encoded in the quantum device, the many-body Rydberg
system acts as a fixed nonlinear feature map, with learning performed
entirely in classical post-processing---conserving both energy and
computational resources relative to deep-learning training pipelines.

\subsection{Motivation for Quantum-Native Fingerprinting}

A fundamental limitation of our earlier SDR work~\cite{sharma2025} was
that image \emph{matching} was performed by a classical Chamfer distance~\cite{borgefors1986}
computed on the post-simulation Rydberg density plots---a classical
geometric metric applied to a quantum output.
This approach failed to exploit the rich many-body quantum correlations
generated by the Rydberg Hamiltonian, and discarded the most quantum
information the device produced.

The present work addresses this gap by extracting a fully quantum
fingerprint directly from the evolved quantum state: the
Pearson-normalized connected correlation matrix and the derived 2D static structure factor.
The structure factor is a well-established condensed-matter
observable~\cite{juliafarreAQM2024} routinely used to characterize
quantum phases in Rydberg arrays; here we apply it as an image
descriptor for the first time.
Matching is then performed by cosine similarity on the
structure-factor fingerprint---a scale-invariant metric appropriate
for Fourier-domain representations.

\subsection{Cartographic Generalization and RDP}

Cartographic generalization (CG)~\cite{ramer1972,douglas1973,song2016}
denotes the systematic simplification of spatial data while preserving
its essential geometric structure.
Traditionally developed within cartography to enable the depiction of
geographic entities---roads, rivers, coastlines---across varying map
scales, CG encompasses line simplification, smoothing, aggregation, and
displacement techniques.
Its computational efficiency, geometric interpretability, and
parameter-based control over the degree of simplification make it
indispensable not only in cartography but also in computer vision,
robotics, and image-based shape abstraction.

The most influential line simplification technique in CG is the
Ramer--Douglas--Peucker (RDP) algorithm~\cite{ramer1972,douglas1973},
independently proposed by Urs Ramer (1972) and later by Douglas
and Peucker (1973).
RDP operates on the principle of recursive distance thresholding.
Given a polyline $\{P_i\}_{i=1}^{n}$, the algorithm computes the
perpendicular distance of each intermediate point $P_k$ from the
baseline $\overline{P_1 P_n}$:
\begin{equation}
  \label{eq:rdp_dist}
  d\!\left(P_k,\,\overline{P_1 P_n}\right)
  = \frac{\left\|\left(P_n - P_1\right)
  \times\left(P_1 - P_k\right)\right\|}
  {\left\|P_n - P_1\right\|}.
\end{equation}
This perpendicular distance is the decision metric of RDP.
A point $P_k$ is recursively retained if and only if
\begin{equation}
  \label{eq:rdp_crit}
  \max_{k \in \{2,\ldots,n-1\}},
  d\!\left(P_k,\,\overline{P_1 P_n}\right) > \varepsilon,
\end{equation}
where $\varepsilon$ is the user-specified tolerance parameter; points
with distance less than $\varepsilon$ are discarded.
By adjusting $\varepsilon$, users balance geometric precision against
data compactness.
In our pipeline, $\varepsilon$ is adaptively raised until the retained
point count satisfies the hardware atom-number cap, guaranteeing a
structurally faithful yet qubit-frugal atom array.

\section{Related Work}
\label{sec:related}

Most prior quantum image processing research operates in the digital
paradigm, employing representations such as FRQI, QPIE, and NEQR whose
qubit requirements scale with image resolution~\cite{su2020}.
Quantum circuit simulations of low-resolution images emphasize
exponential speedups but remain impractical for real-world image sizes.

Yao et al.~\cite{yao2017} demonstrated NMR-based quantum edge
detection on a $4\times 4$ image using QPIE encoding and Quantum
Hadamard Edge Detection (QHED), achieving $\mathcal{O}(1)$ boundary
detection.
Despite the elegance of this result, qubit scarcity and NMR noise
present severe obstacles to scaling.
Cavalieri and Maio~\cite{cavalieri2020} improved QHED using central
finite differences, but validated results through classical simulation
only; decoherence and gate noise degrade the results severely on
physical gate-based hardware.

Kornja\v{c}a et al.~\cite{kornjaca2024} introduced quantum reservoir
computing on QuEra's Aquila by encoding downscaled tomato-leaf images
via vertical inter-atom distances in a $9\times9$ atom grid, yielding
72-pixel resolution with 81 qubits.
This scheme scales qubit count linearly with pixel count: 108 qubits
for 96 pixels.
Such resolutions are insufficient for most real-world applications.

Lu et al.~\cite{lu2024} proposed hybrid digital-analog Rydberg
learning on Quantum Phase (QPL) and MNIST.
Their approach requires trainable parameters, a classical optimiser,
and labelled data---it is a supervised classifier, not an unsupervised
retrieval system.

In contrast, our Sparse-Dots Representation (SDR) overcomes all these
limitations. The qubit count is determined solely by geometric
complexity---governed by the application-specific tolerance parameter
and available hardware resources---rather than by pixel resolution,
making SDR directly compatible with Aquila's global analog control.
To our knowledge, SDR is the only analog quantum image representation
with this property, enabling compact encoding of real-world image sizes
and high-fidelity simulation of large, geometrically complex structures---a flexibility unavailable to techniques
like FRQI~\cite{su2020}, NEQR~\cite{zhang2013} or
QPIE~\cite{yao2017}.
The present work further advances SDR by replacing the classical
Chamfer-distance matching~\cite{sharma2025} with a fully
quantum-native fingerprint, based on many-body correlations and the
2-D static structure factor.

\begin{table*}[t]
  \caption{\label{tab:qimr}%
    Comparison of quantum image-processing methods and classical SIFT\@.
  Image dimension ($2^n\!\times\!2^n$ pixels).}
  \begin{ruledtabular}
    \begin{tabular}{lllll}
      Method & Encoding & Observables & Task & Fingerprint dim \\
      \hline
      \textbf{This work} &
      Sobel $+$ RDP sparse dots &
      $\tilde{C}_{ij},\;S(\bm{k})$ &
      Classification &
      Variable(72 std.) \\
      Kornja\v{c}a 2024~\cite{kornjaca2024} &
      Atom-position/local-detuning &
      $\langle Z_j\rangle,\langle Z_jZ_k\rangle$ &
      Time-series,Classification &
      Variable \\
      Lu 2024~\cite{lu2024} &
      PCA product states &
      Single-qubit &
      Supervised MNIST,QPL &
      N/A \\
      FRQI/QPIE/NEQR~\cite{su2020,zhang2013} &
      Pixel-amplitude/basis-state &
      Amplitude/Basis-state &
      General QImP &
      $4^n$-scale \\
      Classical SIFT &
      Keypoint descriptors &
      Local image gradients &
      Matching/retrieval &
      Variable(128 std.) \\
    \end{tabular}
  \end{ruledtabular}
\end{table*}

\noindent
Table~\ref{tab:qimr} summarizes the comparison.
The key differentiator of this work is: a variable-length
fingerprint enabling direct comparison across images with different
atom counts, with the first use of the static structure factor as an
image descriptor.

\section{Method: Sparse-Dots Representation and Quantum Fingerprinting}
\label{sec:method}

\subsection{Rydberg Hamiltonian}

The dynamics of the neutral-atom array on Aquila are governed by the
time-dependent Rydberg Hamiltonian~\cite{wurtz2023}:
\begin{equation}
  \label{eq:hamiltonian}
  \begin{split}
    H(t) &= \frac{\Omega(t)}{2}\sum_{j}
    \bigl(|g_j\rangle\langle r_j| + |r_j\rangle\langle g_j|\bigr)
    + \sum_{j<k} V_{jk}\,\nn_j\nn_k \\
    &\quad - \sum_{j}
    \bigl[\Delta_g(t) + \alpha_j\Delta_l(t)\bigr]\nn_j,
  \end{split}
\end{equation}
where $\Omega(t)$ is the global Rabi frequency,
$\nn_j=|r_j\rangle\langle r_j|$ is the Rydberg number operator at
site $j$, $\Delta_g(t)$ is the global detuning, $\Delta_l(t)$ is an
optional local detuning scaled by the site factor $\alpha_j$, and
\begin{equation}
  \label{eq:vdw}
  V_{jk} = \frac{C_6}{|\rvec_j - \rvec_k|^6}
\end{equation}
is the van der Waals blockade interaction.
This $r^{-6}$ dependence is the key mechanism: atoms placed according
to an image's geometric skeleton interact with strengths that precisely
encode the pairwise distances of the image contour, embedding the
image's shape into the many-body quantum state $|\psi(T)\rangle$.
We follow the standard Bloqade/QuEra convention in which the drive term
is written with an explicit factor of $\tfrac{1}{2}$, so that a
resonant $\pi$-pulse satisfies $\Omega_{\max}\,t_\pi = \pi$.

\subsection{Pipeline Overview}

The full Stage~1 workflow (without QRC), illustrated in
Fig.~\ref{fig:flowchart}, consists of a classical pre-processing stage
followed by quantum time-evolution and a quantum-native fingerprinting
and matching stage.
\begin{figure}[htbp]
  \centering
  \scalebox{0.58}{
    \begin{tikzpicture}[
        node distance=0.75cm, auto,
        myarrow/.style={-Stealth, thick, draw=black!70},
        base/.style={rectangle, draw=black!45, thick, align=center,
          text width=6.4cm, minimum height=0.9cm, inner sep=7pt,
        drop shadow, font=\sffamily\large},
        boxPink/.style  ={base, draw=red!60,        fill=red!8},
        boxYellow/.style={base, draw=orange!60,      fill=yellow!8},
        boxGreen/.style ={base, draw=green!60!black, fill=green!8},
        boxBlue/.style  ={base, draw=blue!60,        fill=blue!8},
        boxPurple/.style={base, draw=violet!60,      fill=violet!8},
        boxOrange/.style={base, draw=orange!80!red,  fill=orange!15},
        boxTeal/.style  ={base, draw=teal!70,        fill=teal!8},
        boxCrimson/.style={base, draw=red!80!black,  fill=red!15}
      ]
      \node[boxPink]   (start)   {Read Input Image};
      \node[boxYellow, below=of start]   (sobel)
      {Generate Edges using Sobel Filter};
      \node[boxGreen,  below=of sobel]   (coords)
      {Convert to $(x,y)$ Coordinates and Subsample\\
      (Uniform Stride Pre-filter of Edge Pixels)};
      \node[boxBlue,   below=of coords]  (ramer)
      {Dot Minimization using\\
      Ramer--Douglas--Peucker Algorithm};
      \node[boxPurple, below=of ramer]   (atoms)
      {Convert Dots to Atoms with Micrometre Coordinates};
      \node[boxGreen,  below=of atoms]   (waveforms)
      {Define Global Waveforms: Rabi Drive $\Omega(t)$
      and Detuning $\Delta(t)$};
      \node[boxOrange, below=of waveforms] (hamiltonian)
      {Engineer Rydberg Hamiltonian
      $H(t)$~[Eq.~\eqref{eq:hamiltonian}]};
      \node[boxBlue,   below=of hamiltonian] (bloqade)
      {Time-Evolve $|\psi(T)\rangle$ via Bloqade SDK\\
      (Schr\"{o}dinger Equation Integration)};
      \node[boxTeal,   below=of bloqade]     (pearson)
      {Compute Pearson-Normalized
      Correlation Matrix $\tilde{C}_{ij}$~[Eq.~\eqref{eq:pearson}]};
      \node[boxCrimson,below=of pearson]     (sfactor)
      {Compute 2-D Structure Factor
      $S(\bm{k})$ on $9\!\times\!8$ Grid~[Eq.~\eqref{eq:sfactor}]};
      \node[boxYellow, below=of sfactor]     (cosine)
      {Match via Cosine Similarity
      on 72-Element Fingerprint~[Eq.~\eqref{eq:cosine}]};
      \node[boxPink,   below=of cosine]      (display)
      {Display Best Match with Cosine Similarity Score};
      \draw[myarrow] (start)       -- (sobel);
      \draw[myarrow] (sobel)       -- (coords);
      \draw[myarrow] (coords)      -- (ramer);
      \draw[myarrow] (ramer)       -- (atoms);
      \draw[myarrow] (atoms)       -- (waveforms);
      \draw[myarrow] (waveforms)   -- (hamiltonian);
      \draw[myarrow] (hamiltonian) -- (bloqade);
      \draw[myarrow] (bloqade)     -- (pearson);
      \draw[myarrow] (pearson)     -- (sfactor);
      \draw[myarrow] (sfactor)     -- (cosine);
      \draw[myarrow] (cosine)      -- (display);
    \end{tikzpicture}
  }
  \caption{Complete pipeline for quantum-native image matching (Stage~1).
    Upper eight blocks constitute the classical pre-processing and
    quantum encoding stage introduced in~\cite{sharma2025}.
    Lower four blocks constitute the new quantum fingerprinting
    and matching stage introduced in this work, replacing the
  classical Chamfer-distance matching used previously.}
  \label{fig:flowchart}
\end{figure}

\subsection{Classical Pre-Processing: SDR Construction}

\paragraph{Sobel edge extraction.}
The input image is converted to grayscale and convolved with the Sobel
kernels $S_x$ and $S_y$ to yield gradients
$G_x = S_x \ast I$ and $G_y = S_y \ast I$, whose magnitudes
approximate $|\nabla I| \approx |G_x| + |G_y|$.
Edge pixels are identified as those exceeding a threshold $\tau$
(typically 0.3-0.5 of the normalised magnitude).

\paragraph{Uniform pre-sampling.}
The set of edge pixels is uniformly sub-sampled with stride $d$
(typically 50-90 pixels) to produce a tractable initial point cloud.

\paragraph{RDP with atom cap.}
The RDP algorithm is applied with tolerance $\varepsilon$.
To guarantee that the resulting atom count $N\le N_{\max}$ (typically
24, determined by the computational budget $2^N$), $\varepsilon$ is
adaptively increased by a factor of 1.25 per iteration until the cap is
satisfied.
This eliminates both the duplicate-atom crash (which would produce
$V_{jk}\to\infty$) and the non-geometric uniform-index subsampling of
earlier structure-factor codes.

\paragraph{Physical scaling.}
Each retained $(x,y)$ pixel coordinate is multiplied by the physical
pixel size $\delta_\mu$ (in $\mu$m/pixel) to produce the atom
coordinates $\rvec_j$ passed to Bloqade.

\begin{figure}[htbp]
  \centering
  \newcommand{\fw}{3.5cm}
  \setlength{\tabcolsep}{1em}
  \renewcommand{\arraystretch}{1.2}
  \begin{tabular}{ccc}
    \begin{tabular}[t]{@{}c@{}}
      \includegraphics[width=\fw,height=\fw,keepaspectratio]{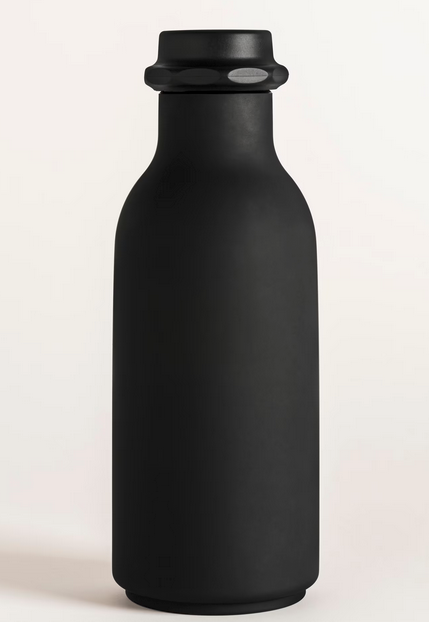}\\
      {\small (a) Original (0.3\,MP)}
    \end{tabular}&
    \textcolor{green!70!black}{\Huge$\rightarrow$}&
    \begin{tabular}[t]{@{}c@{}}
      \includegraphics[width=\fw,height=\fw,keepaspectratio]{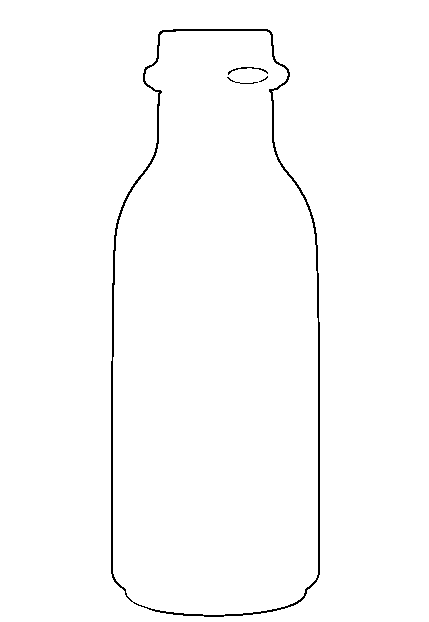}\\
      {\small (b) Sobel edges}
    \end{tabular}\\[0.5em]
    &&\textcolor{green!70!black}{\Huge$\downarrow$}\\[0.3em]
    \begin{tabular}[t]{@{}c@{}}
      \includegraphics[width=\fw,height=\fw,keepaspectratio]{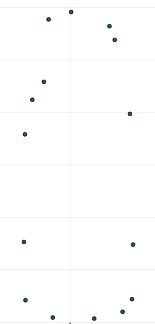}\\
      {\small (d) Sparse-Dots image}
    \end{tabular}&
    \textcolor{green!70!black}{\Huge$\leftarrow$}&
    \begin{tabular}[t]{@{}c@{}}
      \includegraphics[width=\fw,height=\fw,keepaspectratio]{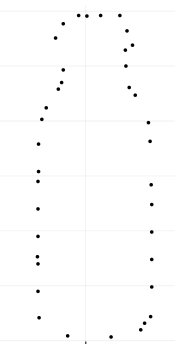}\\
      {\small (c) Dotted version}
    \end{tabular}
  \end{tabular}
  \caption{Image pre-processing steps.
    (a)~Original 0.3-megapixel image.
    (b)~Sobel filter output.
    (c)~Edge pixels replaced by equidistant dots.
    (d)~Sparse-Dots image after RDP compression
  (typically $N=10$--$20$ dots).}
  \label{fig:processing}
\end{figure}

\subsection{Quantum Encoding and Time Evolution}

The $N$ atom positions $\{\rvec_j\}$ are loaded into the programmable
tweezer array.
The quantum register is initialised in the all-ground state
$|\psi(0)\rangle = |g\rangle^{\otimes N}$.
The Rydberg Hamiltonian~[Eq.~(\ref{eq:hamiltonian})] is applied with a
piecewise-linear Rabi drive
\begin{equation*}
  \Omega(t)\,:\quad 0\xrightarrow{0.2\,\mu{\rm s}}\Omega_{\max}
  \xrightarrow{0.6\,\mu{\rm s}[\text{plateau}]}\Omega_{\max}
  \xrightarrow{0.2\,\mu{\rm s}} 0,
\end{equation*}
and a simultaneous three-segment detuning waveform
\begin{equation*}
  \Delta(t)\,:\;
  \begin{cases}
    -2\pi\!\times\!12\,{\rm MHz} & 0 \le t \le 0.2\,\mu{\rm s},\\
    \text{linear ramp to }+2\pi\!\times\!10\,{\rm MHz}
    & 0.2 < t < 0.8\,\mu{\rm s},\\
    +2\pi\!\times\!10\,{\rm MHz} & 0.8 \le t \le 1.0\,\mu{\rm s},
  \end{cases}
\end{equation*}
over total evolution time $T=1\,\mu{\rm s}$.
The detuning sweep drives the system from a disordered regime toward a
blockade-constrained many-body configuration determined by the image geometry.

\section{Quantum Fingerprinting: Correlations and Structure Factor}
\label{sec:fingerprint}

This section presents the new contribution of this paper: the
extraction of a fixed-length, quantum-native image fingerprint from
the evolved state $|\psi(T)\rangle$.
The quantum character of the protocol arises entirely from the
many-body interference encoded in the connected correlators, which
cannot be reconstructed from single-particle observables alone.

\subsection{Exact Statevector Expectation Values}
\label{subsec:sv}

In the simulation, we work with the full statevector
$|\psi\rangle\in\mathbb{C}^{2^N}$.
The single-site Rydberg occupation and the two-site joint occupancy are
computed exactly as
\begin{align}
  \langle\nn_i\rangle &= \sum_{k=0}^{2^N-1}
  |\psi_k|^2\,n_i^{(k)}, \label{eq:pop} \\
  \langle\nn_i\nn_j\rangle &= \sum_{k=0}^{2^N-1}
  |\psi_k|^2\,n_i^{(k)}\,n_j^{(k)}, \label{eq:joint}
\end{align}
where $n_i^{(k)}\in\{0,1\}$ is the Rydberg occupation of atom $i$ in
basis state $k$.
Bloqade/Yao uses LSB-first (little-endian) qubit ordering: qubit $i$
occupies bit position $i-1$ of $k$, i.e.\ $n_i^{(k)} = (k\,\gg\,(i-1))\,\&\,1$.
On hardware, $\langle\nn_i\rangle$ is estimated from projective
measurements (Rydberg density readout), and
$\langle\nn_i\nn_j\rangle$ from shot-based joint excitation counts.

\subsection{Pearson-Normalized Correlation Matrix}
\label{subsec:pearson}

The raw connected two-site correlator is
$C_{ij}^{\rm raw} = \langle\nn_i\nn_j\rangle
- \langle\nn_i\rangle\langle\nn_j\rangle$.
Because $\nn_i$ is a projector ($\nn_i^2 = \nn_i$), its variance is
exactly
$\mathrm{Var}(\nn_i) = \langle\nn_i\rangle(1-\langle\nn_i\rangle)$.
The Pearson-normalized correlation is
\begin{equation}
  \label{eq:pearson}
  \Ctilde_{ij} = \frac{\langle\nn_i\nn_j\rangle
  - \langle\nn_i\rangle\langle\nn_j\rangle}
  {\sqrt{\mathrm{Var}(\nn_i)\cdot\mathrm{Var}(\nn_j)}
  + \varepsilon_{\text{norm}}},
\end{equation}
where $\varepsilon_{\text{norm}}=10^{-8}$ prevents division by zero
when a site has zero variance.
By construction $\Ctilde_{ij}\in[-1,+1]$, with
$\Ctilde_{ij}=-1$ indicating perfect blockade suppression,
$\Ctilde_{ij}=0$ statistical independence, and $\Ctilde_{ij}=+1$
perfect co-excitation.
Importantly, $\tilde{C}_{ij}$ is \emph{amplitude-invariant}: arrays
with identical blockade geometry but different overall Rabi responses
produce identical $\tilde{C}$ matrices.

\subsection{Two-Dimensional Static Structure Factor}
\label{subsec:sfactor}

The 2-D static structure factor is the cosine transform of the
Pearson-normalized correlation matrix over atom-pair displacements:
\begin{equation}
  \label{eq:sfactor}
  S(\kvec) = \frac{1}{N^2}\sum_{i=1}^{N}\sum_{j=1}^{N}
  \Ctilde_{ij}\,\cos\!\bigl[\kvec\cdot(\rvec_i-\rvec_j)\bigr].
\end{equation}
This observable is a condensed-matter standard routinely used to
characterize ordered phases in Rydberg arrays.
Here we repurpose it as an image descriptor.
We evaluate $S(\kvec)$ on the $9\times8$ grid
\begin{equation*}
  k_x\in[-0.30,\,+0.30]\;\text{rad/}\mu\text{m},\quad
  k_y\in[0,\,+0.30]\;\text{rad/}\mu\text{m},
\end{equation*}
exploiting the inversion symmetry $S(k_x,k_y)=S(-k_x,-k_y)$ to
restrict $k_y\ge0$, while retaining both signs of $k_x$ to preserve
sensitivity to left--right orientation asymmetry.
This produces a 72-element fingerprint vector $\bm{S}$ of \emph{fixed
length for all images}, independent of atom count $N$.

\subsection{Cosine Similarity Matching}
\label{subsec:cosine}

Given the fingerprint vectors $\bm{S}_A$ and $\bm{S}_B$ of a query
image $A$ and a database image $B$, the match score is
\begin{equation}
  \label{eq:cosine}
  \mathrm{sim}(A,B) = \cos\theta =
  \frac{\bm{S}_A\cdot\bm{S}_B}{\|\bm{S}_A\|\,\|\bm{S}_B\|}.
\end{equation}
Cosine similarity is preferred over Euclidean distance because it is
invariant to overall scale: if $\bm{S}_B=\lambda\bm{S}_A$ for any
$\lambda>0$, then $\mathrm{sim}=1$ regardless of $\lambda$.
This makes the metric robust to hardware amplitude fluctuations.

\section{Quantum Fingerprinting in Quantum Reservoir Computing}
\label{sec:QRC}

Quantum Reservoir Computing (QRC) is an emerging paradigm in quantum
machine learning that leverages the complex dynamics of quantum systems
to perform efficient information processing.
By utilizing a quantum system as a fixed, high-dimensional dynamical
``reservoir,'' QRC maps input data into a vastly expanded Hilbert
space where nonlinear transformations and memory effects can be
harnessed without training the internal quantum nodes.
Compared to classical machine learning techniques, particularly
recurrent neural networks, QRC offers significant advantages: it
mitigates the vanishing gradient problem, requires training only on a
simple linear readout layer, and can exploit genuine quantum effects
such as superposition and entanglement to achieve superior
computational capacity and expressivity for certain tasks.
It also offers the twin advantages of using much smaller training data
and fewer training cycles.

Here in Stage~2, we extend the quantum fingerprinting approach to Single-step QRC~\cite{liu2026},
as an initial proof-of-concept.
The flowchart Fig.~\ref{fig:flowchart2} depicts all its steps.

\begin{figure}[htbp]
  \centering
  \scalebox{0.58}{
    \begin{tikzpicture}[
        node distance=0.52cm,
        myarrow/.style={-Stealth, thick, draw=black!70},
        mydia/.style={diamond, draw=orange!80!red, thick, fill=orange!10,
          align=center, text width=1.8cm, aspect=2.3,
        inner sep=2pt, drop shadow, font=\sffamily\large},
        bypass/.style={rectangle, draw=green!60!black, thick, fill=green!8,
          align=center, text width=3.0cm, minimum height=0.85cm,
        inner sep=6pt, drop shadow, font=\sffamily\large},
        base/.style={rectangle, draw=black!45, thick,
          align=center, text width=6.4cm,
          minimum height=0.9cm, inner sep=7pt,
        drop shadow, font=\sffamily\large},
        boxPink/.style   ={base, draw=red!60,        fill=red!8},
        boxYellow/.style ={base, draw=orange!60,      fill=yellow!8},
        boxGreen/.style  ={base, draw=green!60!black, fill=green!8},
        boxBlue/.style   ={base, draw=blue!60,        fill=blue!8},
        boxPurple/.style ={base, draw=violet!60,      fill=violet!8},
        boxOrange/.style ={base, draw=orange!80!red,  fill=orange!15},
        boxTeal/.style   ={base, draw=teal!70,        fill=teal!8},
        boxCrimson/.style={base, draw=red!80!black,   fill=red!15},
      ]
      \node[boxPink]   (config)
      {Initialise \texttt{QRCConfig} \& Validate Dataset};
      \node[mydia, below=0.60cm of config] (cacheq)
      {JLD2\\Cache\\Exists?};
      \node[boxOrange, below=0.60cm of cacheq] (extract)
      {Parallel Extraction via \texttt{Threads.@threads}};
      \node[boxYellow, below=of extract] (sobel)
      {Sobel Edges $\to$ Stride Pre-filter $\to$ Adaptive RDP};
      \node[boxGreen,  below=of sobel]   (hamiltonian)
      {Atom Coordinates $\to$ Rydberg Hamiltonian $H(t)$};
      \node[boxTeal,   below=of hamiltonian] (bloqade)
      {Quantum Time Evolution via Bloqade SDK};
      \node[boxCrimson,below=of bloqade]     (pearson)
      {Pearson $\tilde{C}_{ij}$ $\to$
      Structure Factor $S(\bm{k})$};
      \node[boxBlue,   below=of pearson]     (incrsave)
      {Incremental JLD2 Cache Write};
      \node[bypass, right=2.8cm of cacheq]   (loadcache)
      {Load Cache:\\Fingerprints \& Labels};
      \node[boxPurple, below=0.60cm of incrsave] (split)
      {Stratified Train\,/\,Test Split (60\,/\,40)};
      \node[boxOrange, below=of split]   (features)
      {Feature Matrix Assembly
        (\texttt{:flat} $|$ \texttt{:augmented}
      $|$ \texttt{:multiangle})};
      \node[boxBlue,   below=of features] (cv)
      {5-Fold Cross-Validation $\to$ Best $\lambda$};
      \node[boxGreen,  below=of cv]      (ridge)
      {Train Ridge Regression Readout (One-vs-All)};
      \node[boxTeal,   below=of ridge]   (evaluate)
      {Evaluate: Accuracy, $F_1$, Confusion Matrix};
      \node[boxPink,   below=of evaluate] (report)
      {Save Model \& Display Report};
      \draw[myarrow] (config) -- (cacheq);
      \draw[myarrow] (cacheq.south)
      -- node[left,font=\sffamily\large,xshift=-4pt]{\textbf{No}}
      (extract.north);
      \draw[myarrow] (cacheq.east)
      -- node[above,font=\sffamily\large]{\textbf{Yes}}
      (loadcache.west);
      \draw[myarrow] (extract)     -- (sobel);
      \draw[myarrow] (sobel)       -- (hamiltonian);
      \draw[myarrow] (hamiltonian) -- (bloqade);
      \draw[myarrow] (bloqade)     -- (pearson);
      \draw[myarrow] (pearson)     -- (incrsave);
      \draw[myarrow] (incrsave)    -- (split);
      \draw[myarrow] (loadcache.south) |- (split.east);
      \draw[myarrow] (split)    -- (features);
      \draw[myarrow] (features) -- (cv);
      \draw[myarrow] (cv)       -- (ridge);
      \draw[myarrow] (ridge)    -- (evaluate);
      \draw[myarrow] (evaluate) -- (report);
    \end{tikzpicture}
  }
  \caption{QRC pipeline with quantum fingerprints (Stage~2).}
  \label{fig:flowchart2}
\end{figure}

\section{Simulation}
\label{sec:simulation}

The complete pipeline was implemented in Julia~\cite{bezanson2017}
using the Bloqade.jl SDK~\cite{bloqade2023quera}.
All simulations used the global pulse parameters described in
Sec.~\ref{sec:method}.
For Stage~1 (without QRC), a representative database of more than 100
industrial objects was matched against query images using the quantum
structure-factor fingerprint.
Across all tested images, successful matching (correct image ranked
first by cosine similarity) was achieved with 10--21 atoms---well
within the 256-qubit capacity of Aquila.

The quantum-fingerprint approach offers two advantages over the Chamfer
distance used in our earlier work~\cite{sharma2025}: (i)~it exploits
the full many-body quantum state rather than a classical point-cloud
post-processing of single-site occupancies, and (ii)~the 72-element
fingerprint is fixed-length and directly comparable across images with
different atom counts. The Stage 1 results are shown in Fig.~\ref{fig:summary}. The images for stage 1 were taken from "www.pexels.com" under CC0 (Creative Commons Zero) license.

\begin{figure*}[!t]
  \centering
  \begin{minipage}[t]{0.57\linewidth}
    \centering
    \captionof{table}{Cosine similarity between the query
      \texttt{Ball.jpg} and each database image for three values of the
      edge-point stride (\texttt{dot\_spacing}).
      Colour intensity: green = high, red = low.
    The correct match always scores 1.000000.}
    \label{tab:similarities}
    \vspace{4pt}
    \renewcommand{\arraystretch}{1.2}
    \small\setlength{\tabcolsep}{3pt}
    \begin{tabular}{lccc}
      \toprule
      \textbf{Image} & \textbf{70\,px} & \textbf{60\,px} & \textbf{50\,px}\\
      \midrule
      Ball        & \similaritycell{1.000000} & \similaritycell{1.000000} & \similaritycell{1.000000}\\
      Ball\_1     & \similaritycell{0.996364} & \similaritycell{0.988918} & \similaritycell{0.816494}\\
      Ball\_2     & \similaritycell{0.996522} & \similaritycell{0.988927} & \similaritycell{0.926619}\\
      Book        & \similaritycell{0.995900} & \similaritycell{0.905159} & \similaritycell{0.782584}\\
      Bag         & \similaritycell{0.932158} & \similaritycell{0.908217} & \similaritycell{0.934070}\\
      Phone\_2    & \similaritycell{0.833983} & \similaritycell{0.960598} & \similaritycell{0.868721}\\
      Chair\_1    & \similaritycell{0.686997} & \similaritycell{0.734370} & \similaritycell{0.789744}\\
      Chair\_2    & \similaritycell{0.817985} & \similaritycell{0.847800} & \similaritycell{0.603677}\\
      Clothesclip & \similaritycell{0.995949} & \similaritycell{0.801436} & \similaritycell{0.656725}\\
      Bottle      & \similaritycell{0.821039} & \similaritycell{0.794911} & \similaritycell{0.706785}\\
      \bottomrule
    \end{tabular}
  \end{minipage}
  \hfill
  \begin{minipage}[t]{0.39\linewidth}
    \centering
    \captionof{table}{Quantum fingerprint diagnostics for the run with
      \texttt{dot\_spacing} = 70\,px: number of atoms $N$, maximum
      off-diagonal Pearson correlation $\max|\tilde{C}_{ij}|$, and
      Euclidean norm $\|\mathbf{S}\|_2$.
      The $\checkmark$ marks atom pairs with $\max|\tilde{C}_{ij}|=1.0$,
    indicating perfect blockade correlation.}
    \label{tab:diagnostics}
    \vspace{4pt}
    \renewcommand{\arraystretch}{1.2}
    \scriptsize\setlength{\tabcolsep}{2pt}
    \begin{tabular}{lcccc}
      \toprule
      \textbf{Image} & $N$ & $\max|\tilde{C}_{ij}|$ & $\|\mathbf{S}\|_2$
      & $=\!1$?\\
      \midrule
      Ball        & 18 & 0.4551 & 0.3019 & \\
      Ball\_1     & 16 & 0.0365 & 0.3306 & \\
      Ball\_2     & 11 & 0.0254 & 0.4816 & \\
      Book        & 14 & 0.1001 & 0.3759 & \\
      Bag         & 16 & 1.0000 & 0.3444 & \checkmark\\
      Phone\_2    & 20 & 1.0000 & 0.3205 & \checkmark\\
      Chair\_1    & 19 & 1.0000 & 0.4029 & \checkmark\\
      Chair\_2    & 20 & 1.0000 & 0.3265 & \checkmark\\
      Clothesclip & 10 & 0.0582 & 0.5284 & \\
      Bottle      & 20 & 1.0000 & 0.4350 & \checkmark\\
      \bottomrule
    \end{tabular}
  \end{minipage}
  \captionof{figure}{Stage~1 results.
    (a)~Cosine similarities confirm perfect self-match and robust
    ordering across three \texttt{dot\_spacing} values.
  (b)~Diagnostic values for \texttt{dot\_spacing}$=(70,60,50$)\,px.}
  \label{fig:summary}
\end{figure*}

For Stage~2 (QRC), the quantum fingerprints were used to train and
test a 5-class classifier (Dumbell, Mouse Pad, Ottoman, Sofa, Table)
with 50 images per class. The datasets were taken from Amazon-Berkeley objects database, under the "CC BY 4.0" License~\cite{collins2022abo}. The images were rotated by small angles to generate 50 images per class with different view-angles.
Since QRC requires very little labelled data, this small-scale
experiment is intended as a proof-of-concept; the dataset will be
scaled to 300 images per class in future work.
Additionally, Sparse-Dots images can be synthetically generated seamlessly (by varying RDP tuning).

The Stage~2 results are presented in
Tables~\ref{tab:qrc_runs}--\ref{tab:qrc_f1heatmap} and
Figs.~\ref{fig:qrc_global}--\ref{fig:qrc_pareto}.

\textbf{Accuracy and efficiency.}
Across all 10 runs the QRC classifier comfortably exceeds the 20\%
random-chance baseline on a 5-class problem.
Run~R1 (Sobel Threshold $\theta{=}0.50$, $d_s{=}90\,\mu$m, $N_{\max}{=}20$) achieves
the highest Top-1 test accuracy of 72.5\% and Macro-$F_1$ of 0.711 in
just 0.8\,min of GPU wall time, making it the best configuration
overall.
The best full-emulation run, R5 ($\theta{=}0.30$, $d_s{=}50\,\mu$m,
$N_{\max}{=}24$), reaches 69.0\% Top-1 and Macro-$F_1{=}0.685$
at a cost of 73.6\,min.
Crucially, Top-3 accuracy is uniformly high across all configurations
(0.91--0.95), meaning the correct class is ranked within the top three
candidates even when it is not ranked first---a practically useful
property for retrieval pipelines that present a short-list to an
operator.
The 5-fold cross-validation accuracy tracks Top-1 closely for
fast-emulation runs (gap of 3-8 pp.) but slightly overestimates
Top-1 in the full-emulation family, reflecting the small dataset size
of 250 images total; scaling to 300 images per class is expected to
close this gap.

\textbf{Parameter sensitivity and class-level behaviour.}
The single most impactful hyperparameter is dot spacing: reducing
$d_s$ from 90 to 70\,$\mu$m (R1$\to$R2) drops Top-1 by 14.3\,pp,
because sparser edge sampling discards discriminative geometric detail
before quantum encoding.
Increasing $N_{\max}$ from 20 to 24 atoms (R4$\to$R5) recovers
$+5$\,pp by enabling richer many-body correlations in the 72-element
structure-factor fingerprint.
Enlarging the fingerprint dimension from 72 to 81 k-points
(R5$\to$R6) conversely \emph{reduces} Top-1 by 9\,pp, consistent with
the additional high-frequency wavevectors capturing noise rather than
discriminative signal at the length scales of 24-atom arrays.
At the per-class level (Table~\ref{tab:qrc_f1heatmap}), Table objects
are the easiest to classify under fast emulation ($F_1{=}0.92$, R1)
owing to their distinctive straight-edge geometry, while Sofa is the
most persistently difficult class ($F_1{\leq}0.69$ across all runs)
due to geometric ambiguity with Ottoman at low atom counts.
Dumbell recognition improves substantially under full emulation
(R5: $F_1{=}0.77$ vs.\ R1: $F_1{=}0.58$), suggesting that the
bilateral symmetry of dumbbells is more faithfully encoded by the
denser, longer-range correlations available at $\theta{=}0.30$.
The system used for the above stages has the following specifications-- CPU-- AMD Ryzen 9 (5900X, 12 cores/24 Threads), GPU -- NVIDIA Geforce RTX 3060 (12 GB), 32 GB DDR4 RAM.


\begin{table*}[htbp]
  \caption{\label{tab:qrc_runs}%
    Full parameter matrix and classification metrics for all 10 QRC
    simulation runs on the 5-class object dataset (Dumbell, Mouse Pad,
      Ottoman, Sofa, Table; 50 images per class; 60/40 stratified
      train/test split; ridge readout with 5-fold CV $\lambda$ selection;
    GPU emulation v3 on NVIDIA RTX~3060).
    \colorbox{QOrange!22}{Orange rows}: Sobel Threshold $\theta{=}0.50$ fast-emulation
    (R1--R3, wall time $<\!2$\,min).
    \colorbox{QBlue!15}{Blue rows}: $\theta{=}0.30$ full-emulation
    (R4--R10).
    $d_s$ = dot spacing (\textmu m/px); $a$ = pixel size
    (\textmu m/px); $N_{\max}$ = max atoms; $d_{\rm FP}$ = fingerprint
    dimension; $T$ = evolution time; $\lambda^*$ = best ridge
    regularisation; $\hat{A}_{\rm CV}$ = 5-fold CV accuracy;
    $A_1$/$A_3$ = Top-1/Top-3 test accuracy;
  $\bar{F}_1$ = Macro-$F_1$.}
  \small\renewcommand{\arraystretch}{1.30}
  \begin{tabular*}{\textwidth}{@{\extracolsep{\fill}}
    c c c c c c c c c c c c r}
    \toprule
    Run & $\theta$ & $d_s$ & $a$ & $N_{\max}$ & $d_{\rm FP}$ &
    $T\,[\mu\text{s}]$ & $\lambda^*$ & $\hat{A}_{\rm CV}$ & $A_1$ &
    $A_3$ & $\bar{F}_1$ & Wall\\
    \midrule
    \rowcolor{QOrange!18}
    R1  & 0.50 & 90 & 1.00 & 20 & 72 & 1.00 & $10^{-3}$ &
    0.692 & \textbf{0.725} & \textbf{0.950} & \textbf{0.711} & 0.8\,min\\
    \rowcolor{QOrange!18}
    R2  & 0.50 & 70 & 1.00 & 20 & 72 & 1.00 & $10^{-2}$ &
    0.546 & 0.582 & 0.888 & 0.575 & 1.1\,min\\
    \rowcolor{QOrange!18}
    R3  & 0.50 & 90 & 1.20 & 20 & 72 & 1.00 & $10^{-3}$ &
    0.563 & 0.638 & 0.938 & 0.633 & 0.8\,min\\
    \rowcolor{QBlue!12}
    R4  & 0.30 & 50 & 1.00 & 20 & 72 & 1.00 & $10^{-1}$ &
    0.651 & 0.640 & 0.940 & 0.634 & 4.6\,min\\
    \rowcolor{QBlue!12}
    R5  & 0.30 & 50 & 1.00 & 24 & 72 & 1.00 & $10^{-1}$ &
    0.665 & 0.690 & 0.940 & 0.685 & 73.6\,min\\
    \rowcolor{QBlue!12}
    R6  & 0.30 & 50 & 1.00 & 24 & \underline{81} & 1.00 & $10^{-1}$ &
    0.672 & 0.600 & 0.920 & 0.587 & 73.4\,min\\
    \rowcolor{QBlue!12}
    R7  & 0.30 & 50 & \underline{0.80} & 24 & 72 & 1.00 & $10^{-1}$ &
    0.658 & 0.640 & 0.940 & 0.621 & 123.0\,min\\
    \rowcolor{QBlue!12}
    R8  & 0.30 & 50 & 1.00 & 24 & 72 & \underline{1.10} & $10^{-1}$ &
    0.658 & 0.680 & 0.950 & 0.673 & 83.4\,min\\
    \rowcolor{QBlue!12}
    R9  & 0.30 & 50 & 0.95 & 24 & 72 & 1.00 & $10^{-1}$ &
    0.672 & 0.660 & 0.940 & 0.647 & 80.0\,min\\
    \rowcolor{QBlue!12}
    R10 & 0.30 & 50 & 1.05 & 24 & 72 & 1.05 & $10^{-1}$ &
    0.591 & 0.580 & 0.910 & 0.573 & 74.0\,min\\
    \midrule
    \multicolumn{8}{l}{\footnotesize Chance level (1/5 random baseline)}
    & 0.200 & 0.200 & 0.600 & 0.200 & ---\\
    \bottomrule
  \end{tabular*}
\end{table*}

\begin{table*}[htbp]
  \caption{\label{tab:qrc_f1heatmap}%
    Per-class $F_1$ score heatmap for all 10 QRC runs.
    Colour grade:
    \colorbox{HF0!50}{\;$<$0.40\;} poor,
    \colorbox{HF1!50}{\;0.40--0.55\;},
    \colorbox{HF2!80}{\;0.55--0.70\;},
    \colorbox{HF3!60}{\;0.70--0.80\;},
    \colorbox{HF4!55}{\;$>$0.80\;} excellent.
    Sofa and Mouse Pad are persistently the hardest classes across all
    runs; Table peaks at $F_1{=}0.92$ in R1 due to its distinctive
  straight-edge geometry at high dot spacing.}
  \small\renewcommand{\arraystretch}{1.45}
  \begin{tabular*}{\textwidth}{@{\extracolsep{\fill}} l ccccc c}
    \toprule
    \textbf{Run} & \textbf{Dumbell} & \textbf{Mouse Pad} &
    \textbf{Ottoman} & \textbf{Sofa} & \textbf{Table} &
    Macro-$F_1$\\
    \midrule
    R1  & \fclr{0.583}&\fclr{0.692}&\fclr{0.667}&\fclr{0.688}&\fclr{0.923}&\cellcolor{HF3!60}0.711\\
    R2  & \fclr{0.516}&\fclr{0.727}&\fclr{0.523}&\fclr{0.286}&\fclr{0.821}&\cellcolor{HF2!80}0.575\\
    R3  & \fclr{0.444}&\fclr{0.759}&\fclr{0.650}&\fclr{0.667}&\fclr{0.647}&\cellcolor{HF2!80}0.633\\
    \midrule
    R4  & \fclr{0.739}&\fclr{0.550}&\fclr{0.632}&\fclr{0.581}&\fclr{0.667}&\cellcolor{HF2!80}0.634\\
    R5  & \fclr{0.766}&\fclr{0.595}&\fclr{0.778}&\fclr{0.606}&\fclr{0.681}&\cellcolor{HF2!80}0.685\\
    R6  & \fclr{0.636}&\fclr{0.546}&\fclr{0.818}&\fclr{0.270}&\fclr{0.667}&\cellcolor{HF2!80}0.587\\
    R7  & \fclr{0.694}&\fclr{0.424}&\fclr{0.698}&\fclr{0.516}&\fclr{0.773}&\cellcolor{HF2!80}0.621\\
    R8  & \fclr{0.766}&\fclr{0.595}&\fclr{0.732}&\fclr{0.606}&\fclr{0.667}&\cellcolor{HF2!80}0.673\\
    R9  & \fclr{0.625}&\fclr{0.611}&\fclr{0.773}&\fclr{0.500}&\fclr{0.727}&\cellcolor{HF2!80}0.647\\
    R10 & \fclr{0.667}&\fclr{0.444}&\fclr{0.632}&\fclr{0.487}&\fclr{0.634}&\cellcolor{HF2!80}0.573\\
    \bottomrule
  \end{tabular*}
\end{table*}

\begin{figure*}[!t]
  \centering
  \begin{subfigure}[t]{0.48\textwidth}
    \begin{tikzpicture}
      \begin{axis}[
          qrc, title={\textbf{(a)} Top-1 \& CV accuracy},
          ybar=1pt, bar width=7pt,
          width=\linewidth, height=5.8cm,
          xtick={1,...,10},
          xticklabels={R1,R2,R3,R4,R5,R6,R7,R8,R9,R10},
          xticklabel style={font=\scriptsize},
          ymin=0, ymax=1.0, ytick={0,0.2,0.4,0.6,0.8,1.0},
          ylabel={Accuracy}, enlarge x limits=0.07,
          legend pos=north east, legend style={font=\tiny},
        ]
        \addplot[fill=QOrange!80,draw=QOrange!60!black] coordinates
        {(1,0.725)(2,0.582)(3,0.638)};
        \addplot[fill=QBlue!75,draw=QBlue!50!black] coordinates
        {(4,0.640)(5,0.690)(6,0.600)(7,0.640)(8,0.680)(9,0.660)(10,0.580)};
        \addplot[only marks,mark=diamond*,mark size=3pt,
        mark options={fill=QGold,draw=QGold!60!black}] coordinates
        {(1,0.692)(2,0.546)(3,0.563)(4,0.651)(5,0.665)(6,0.672)
        (7,0.658)(8,0.658)(9,0.672)(10,0.591)};
        \addplot[dashed,thick,color=QRed,domain=0.5:10.5,samples=2]{0.200};
        \legend{Top-1 ($\theta{=}0.50$),Top-1 ($\theta{=}0.30$),CV acc,Chance}
      \end{axis}
    \end{tikzpicture}
  \end{subfigure}
  \hfill
  \begin{subfigure}[t]{0.48\textwidth}
    \begin{tikzpicture}
      \begin{axis}[
          qrc, title={\textbf{(b)} Top-3 accuracy \& Macro-$F_1$},
          ybar=1pt, bar width=7pt,
          width=\linewidth, height=5.8cm,
          xtick={1,...,10},
          xticklabels={R1,R2,R3,R4,R5,R6,R7,R8,R9,R10},
          xticklabel style={font=\scriptsize},
          ymin=0, ymax=1.05, ytick={0,0.2,0.4,0.6,0.8,1.0},
          ylabel={Score}, enlarge x limits=0.07,
          legend pos=south east, legend style={font=\tiny},
        ]
        \addplot[fill=QTeal!80,draw=QTeal!60!black] coordinates
        {(1,0.950)(2,0.888)(3,0.938)(4,0.940)(5,0.940)
        (6,0.920)(7,0.940)(8,0.950)(9,0.940)(10,0.910)};
        \addplot[only marks,mark=*,mark size=3pt,
        mark options={fill=QPurple,draw=QPurple!60!black}] coordinates
        {(1,0.711)(2,0.575)(3,0.633)(4,0.634)(5,0.685)
        (6,0.587)(7,0.621)(8,0.673)(9,0.647)(10,0.573)};
        \addplot[dashed,thick,color=QRed,domain=0.5:10.5,samples=2]{0.600};
        \legend{Top-3 acc,Macro-$F_1$,Top-3 chance}
      \end{axis}
    \end{tikzpicture}
  \end{subfigure}
  \caption{Global classification performance for all 10 QRC runs.
    \textbf{(a)}~Top-1 test accuracy (bars) and 5-fold CV accuracy
    (diamonds). Orange = $\theta{=}0.50$ fast-emulation (R1--R3);
    blue = $\theta{=}0.30$ full-emulation (R4--R10).
    R1 achieves the highest Top-1 (72.5\%) and Macro-$F_1$ (0.711) in
    just 0.8\,min; R5 is the best full-emulation run (69.0\%).
    \textbf{(b)}~Top-3 accuracy (bars) and Macro-$F_1$ (circles).
    Top-3 is uniformly high (0.91--0.95) across all runs, confirming
    that correct classes are consistently ranked within the top~3.
    Dashed red lines = random chance levels (20\% for Top-1, 60\%
  for Top-3).}
  \label{fig:qrc_global}
\end{figure*}

\begin{figure*}[!t]
  \centering
  \begin{subfigure}[t]{0.46\textwidth}
    \centering
    {\small\bfseries (a) Run R1 \;|\; Top-1 = 72.5\%}\\[2pt]
    {\footnotesize $\theta{=}0.50$, $d_s{=}90\,\mu$m,
    $a{=}1.00\,\mu$m/px, $N_{\max}{=}20$}\\[4pt]
    \small\setlength{\tabcolsep}{3pt}\renewcommand{\arraystretch}{1.75}
    \begin{tabular}{c|ccccc}
      & Du & MP & Ot & So & Ta\\
      \hline
      Du &\cellcolor{QGreen!47}\textbf{7} &\cellcolor{QRed!7}1
      &\cellcolor{QRed!13}2 &\cellcolor{QRed!27}4 &\cellcolor{QRed!7}1\\
      MP &\cellcolor{QRed!0}0 &\cellcolor{QGreen!60}\textbf{9}
      &\cellcolor{QRed!40}6 &\cellcolor{QRed!0}0 &\cellcolor{QRed!0}0\\
      Ot &\cellcolor{QRed!0}0 &\cellcolor{QRed!0}0
      &\cellcolor{QGreen!93}{\color{white}\textbf{13}}
      &\cellcolor{QRed!7}1 &\cellcolor{QRed!0}0\\
      So &\cellcolor{QRed!6}1 &\cellcolor{QRed!6}1
      &\cellcolor{QRed!19}3 &\cellcolor{QGreen!69}\textbf{11}
      &\cellcolor{QRed!0}0\\
      Ta &\cellcolor{QRed!5}1 &\cellcolor{QRed!0}0
      &\cellcolor{QRed!5}1 &\cellcolor{QRed!0}0
      &\cellcolor{QGreen!90}{\color{white}\textbf{18}}\\
    \end{tabular}\\[3pt]
    {\footnotesize Rows = true class;\; Cols = predicted}
  \end{subfigure}
  \hfill
  \begin{subfigure}[t]{0.46\textwidth}
    \centering
    {\small\bfseries (b) Run R5 \;|\; Top-1 = 69.0\%}\\[2pt]
    {\footnotesize $\theta{=}0.30$, $d_s{=}50\,\mu$m,
    $a{=}1.00\,\mu$m/px, $N_{\max}{=}24$}\\[4pt]
    \small\setlength{\tabcolsep}{3pt}\renewcommand{\arraystretch}{1.75}
    \begin{tabular}{c|ccccc}
      & Du & MP & Ot & So & Ta\\
      \hline
      Du &\cellcolor{QGreen!90}{\color{white}\textbf{18}}
      &\cellcolor{QRed!10}2 &\cellcolor{QRed!0}0
      &\cellcolor{QRed!0}0 &\cellcolor{QRed!0}0\\
      MP &\cellcolor{QRed!35}7 &\cellcolor{QGreen!55}\textbf{11}
      &\cellcolor{QRed!0}0 &\cellcolor{QRed!0}0
      &\cellcolor{QRed!10}2\\
      Ot &\cellcolor{QRed!0}0 &\cellcolor{QRed!0}0
      &\cellcolor{QGreen!70}\textbf{14} &\cellcolor{QRed!5}1
      &\cellcolor{QRed!25}5\\
      So &\cellcolor{QRed!10}2 &\cellcolor{QRed!10}2
      &\cellcolor{QRed!10}2 &\cellcolor{QGreen!50}\textbf{10}
      &\cellcolor{QRed!20}4\\
      Ta &\cellcolor{QRed!0}0 &\cellcolor{QRed!10}2
      &\cellcolor{QRed!0}0 &\cellcolor{QRed!10}2
      &\cellcolor{QGreen!80}\textbf{16}\\
    \end{tabular}\\[3pt]
    {\footnotesize Rows = true class;\; Cols = predicted}
  \end{subfigure}
  \caption{Row-normalised confusion matrices for the two best runs.
    Green diagonal = true-positive fraction;
    red off-diagonal = confusion fraction;
    intensity $\propto$ row-normalised count.
    Abbreviations: Du=Dumbell, MP=Mouse Pad, Ot=Ottoman, So=Sofa,
    Ta=Table.
    \textbf{(a)}~R1 (best overall): Ottoman and Table are nearly
    perfectly classified; Mouse Pad is the dominant confuser, mostly
    misclassified as Ottoman.
    \textbf{(b)}~R5 (best full-emulation): Dumbell achieves 90\%
    recall; Sofa remains the hardest class (50\%), partially
    misclassified as Ottoman and Table due to geometric ambiguity at
  $\leq$24 atoms.}
  \label{fig:qrc_confusion}
\end{figure*}

\begin{figure*}[!t]
  \centering
  \begin{tikzpicture}
    \begin{axis}[
        qrc,
        title={\textbf{Ridge Regularisation Cross-Validation Sweep
        (all 10 runs)}},
        width=\linewidth, height=7.5cm,
        xmode=log,
        xlabel={Regularisation strength $\lambda$},
        ylabel={5-fold CV accuracy},
        ymin=0.18, ymax=0.78,
        xmin=3e-7, xmax=3e2,
        ytick={0.2,0.3,0.4,0.5,0.6,0.7},
        xtick={1e-6,1e-5,1e-4,1e-3,1e-2,1e-1,1e0,1e1,1e2},
        legend columns=2,
        legend pos=south west,
        legend style={font=\tiny, cells={anchor=west},
          column sep=4pt, row sep=-1pt,
          fill=white, fill opacity=0.85,
        draw=gray!40},
      ]
      \addplot[color=QOrange,thick,mark=*,mark size=2pt] coordinates {
        (1e-6,0.3612)(1e-5,0.4490)(1e-4,0.5786)(1e-3,0.6920)
      (1e-2,0.5929)(1e-1,0.3961)(1e0,0.3790)(1e1,0.2627)(1e2,0.2536)};
      \addlegendentry{R1 [$d_s$=90, $\theta$=0.50]}
      \addplot[only marks,mark=star,mark size=5pt,color=QOrange]
      coordinates {(1e-3,0.6920)};
      \addplot[color=QGold,thick,mark=square*,mark size=2pt] coordinates {
        (1e-6,0.3885)(1e-5,0.3818)(1e-4,0.4033)(1e-3,0.4573)
      (1e-2,0.5457)(1e-1,0.5064)(1e0,0.3764)(1e1,0.3354)(1e2,0.3006)};
      \addlegendentry{R2 [$d_s$=70, $\theta$=0.50]}
      \addplot[only marks,mark=star,mark size=5pt,color=QGold]
      coordinates {(1e-2,0.5457)};
      \addplot[color=QRed!70,thick,mark=triangle*,mark size=2pt] coordinates {
        (1e-6,0.3296)(1e-5,0.3695)(1e-4,0.4870)(1e-3,0.5630)
      (1e-2,0.4995)(1e-1,0.3887)(1e0,0.3877)(1e1,0.2536)(1e2,0.2536)};
      \addlegendentry{R3 [$a$=1.20, $\theta$=0.50]}
      \addplot[only marks,mark=star,mark size=5pt,color=QRed!70]
      coordinates {(1e-3,0.5630)};
      \addplot[color=QBlue!55,thick,mark=diamond*,mark size=2pt,dashed]
      coordinates {
        (1e-6,0.4294)(1e-5,0.4294)(1e-4,0.4294)(1e-3,0.4363)
      (1e-2,0.4901)(1e-1,0.6510)(1e0,0.6441)(1e1,0.5370)(1e2,0.4977)};
      \addlegendentry{R4 [$N_{\max}$=20, $\theta$=0.30]}
      \addplot[only marks,mark=star,mark size=5pt,color=QBlue!55]
      coordinates {(1e-1,0.6510)};
      \addplot[color=QBlue,very thick,mark=*,mark size=2.5pt] coordinates {
        (1e-6,0.4237)(1e-5,0.4237)(1e-4,0.4237)(1e-3,0.4370)
      (1e-2,0.5510)(1e-1,0.6651)(1e0,0.5908)(1e1,0.5306)(1e2,0.4101)};
      \addlegendentry{R5 [$N_{\max}$=24, baseline]}
      \addplot[only marks,mark=star,mark size=5pt,color=QBlue]
      coordinates {(1e-1,0.6651)};
      \addplot[color=QPurple,thick,mark=square*,mark size=2pt,dashed]
      coordinates {
        (1e-6,0.5303)(1e-5,0.5303)(1e-4,0.5370)(1e-3,0.5372)
      (1e-2,0.6179)(1e-1,0.6720)(1e0,0.6448)(1e1,0.5777)(1e2,0.5313)};
      \addlegendentry{R6 [$d_{\rm FP}$=81]}
      \addplot[only marks,mark=star,mark size=5pt,color=QPurple]
      coordinates {(1e-1,0.6720)};
      \addplot[color=QCyan!80!black,thick,mark=triangle*,mark size=2pt,dashed]
      coordinates {
        (1e-6,0.5306)(1e-5,0.5306)(1e-4,0.5372)(1e-3,0.5437)
      (1e-2,0.5910)(1e-1,0.6584)(1e0,0.6051)(1e1,0.5244)(1e2,0.4166)};
      \addlegendentry{R7 [$a$=0.80\,\textmu m]}
      \addplot[only marks,mark=star,mark size=5pt,color=QCyan!80!black]
      coordinates {(1e-1,0.6584)};
      \addplot[color=QTeal,thick,mark=pentagon*,mark size=2pt] coordinates {
        (1e-6,0.4306)(1e-5,0.4306)(1e-4,0.4306)(1e-3,0.4439)
      (1e-2,0.5579)(1e-1,0.6584)(1e0,0.5910)(1e1,0.5241)(1e2,0.3699)};
      \addlegendentry{R8 [$T_{\max}$=1.10\,\textmu s]}
      \addplot[only marks,mark=star,mark size=5pt,color=QTeal]
      coordinates {(1e-1,0.6584)};
      \addplot[color=QGreen!75!black,thick,mark=o,mark size=2.5pt,dashed]
      coordinates {
        (1e-6,0.6044)(1e-5,0.6044)(1e-4,0.6044)(1e-3,0.6110)
      (1e-2,0.6246)(1e-1,0.6717)(1e0,0.6110)(1e1,0.5510)(1e2,0.4439)};
      \addlegendentry{R9 [$a$=0.95\,\textmu m]}
      \addplot[only marks,mark=star,mark size=5pt,color=QGreen!75!black]
      coordinates {(1e-1,0.6717)};
      \addplot[color=QViolet,thick,mark=x,mark size=3pt] coordinates {
        (1e-6,0.3885)(1e-5,0.3885)(1e-4,0.3954)(1e-3,0.4021)
      (1e-2,0.5032)(1e-1,0.5910)(1e0,0.5641)(1e1,0.5106)(1e2,0.4099)};
      \addlegendentry{R10 [$a$=1.05, $T$=1.05\,\textmu s]}
      \addplot[only marks,mark=star,mark size=5pt,color=QViolet]
      coordinates {(1e-1,0.5910)};
      \addplot[dotted,thick,color=QRed!70,domain=3e-7:3e2,samples=2]{0.200};
      \addlegendentry{Chance (0.20)}
    \end{axis}
  \end{tikzpicture}
  \caption{5-fold CV accuracy vs.\ ridge regularisation $\lambda$ for
    all 10 runs.
    Stars ($\star$) mark the selected $\lambda^*$ per run.
    Fast-emulation runs (R1--R3, warm tones) peak sharply at
    $\lambda^*{=}10^{-3}$--$10^{-2}$, reflecting compact feature
    spaces generated at high dot spacing.
    Full-emulation runs (R4--R10, cool tones) peak broadly at
    $\lambda^*{=}10^{-1}$, indicating that richer many-body feature
    spaces benefit from stronger regularisation.
    R9 shows a notably flat plateau near $\lambda{=}10^{-2}$--$10^{-1}$,
  consistent with the high-quality features generated at $a{=}0.95$\,\textmu m.}
  \label{fig:qrc_lambda}
\end{figure*}

\begin{figure*}[!t]
  \centering
  \begin{tikzpicture}
    \begin{groupplot}[
        group style={group size=3 by 2,
        horizontal sep=2.0cm, vertical sep=2.5cm},
        qrc,
        width=5.6cm, height=5.0cm,
        ymin=0.45, ymax=0.80,
        ytick={0.50,0.55,0.60,0.65,0.70,0.75,0.80},
        yticklabel style={font=\tiny},
        tick label style={font=\tiny},
        label style={font=\scriptsize\bfseries},
        title style={font=\scriptsize\bfseries},
      ]
      \nextgroupplot[
        title={\textbf{(a)} Dot spacing $d_s$},
        xlabel={$d_s$ [\textmu m/px]},
        ylabel={Top-1 accuracy},
        xtick={70,90}, xmin=62,xmax=98,
        ybar, bar width=18pt,
        nodes near coords,
        nodes near coords style={font=\tiny,yshift=2pt},
      ]
      \addplot[fill=QOrange!80,draw=QOrange!60!black]
      coordinates {(70,0.582)(90,0.725)};
      \addplot[dashed,thick,color=QGreen!70!black,
      domain=62:98,samples=2]{0.690};

      \nextgroupplot[
        title={\textbf{(b)} Pixel size $a$},
        xlabel={$a$ [\textmu m/px]},
        xtick={0.80,0.95,1.00,1.05}, xmin=0.72,xmax=1.13,
        xticklabel style={rotate=30,anchor=north east,font=\tiny},
        ybar, bar width=9pt,
        nodes near coords,
        nodes near coords style={font=\tiny,yshift=2pt},
      ]
      \addplot[fill=QBlue!75,draw=QBlue!60!black]
      coordinates {(0.80,0.640)(0.95,0.660)(1.00,0.690)(1.05,0.580)};
      \addplot[dashed,thick,color=QGreen!70!black,
      domain=0.72:1.13,samples=2]{0.690};

      \nextgroupplot[
        title={\textbf{(c)} Max atoms $N_{\max}$},
        xlabel={$N_{\max}$},
        xtick={20,24}, xmin=18,xmax=26,
        ybar, bar width=20pt,
        nodes near coords,
        nodes near coords style={font=\tiny,yshift=2pt},
      ]
      \addplot[fill=QTeal!75,draw=QTeal!60!black]
      coordinates {(20,0.640)(24,0.690)};
      \addplot[dashed,thick,color=QGreen!70!black,
      domain=18:26,samples=2]{0.690};

      \nextgroupplot[
        title={\textbf{(d)} Fingerprint dim $d_{\rm FP}$},
        xlabel={$d_{\rm FP}$},
        ylabel={Top-1 accuracy},
        xtick={72,81}, xmin=68,xmax=85,
        ybar, bar width=22pt,
        nodes near coords,
        nodes near coords style={font=\tiny,yshift=2pt},
      ]
      \addplot[fill=QPurple!70,draw=QPurple!60!black]
      coordinates {(72,0.690)(81,0.600)};
      \addplot[dashed,thick,color=QGreen!70!black,
      domain=68:85,samples=2]{0.690};

      \nextgroupplot[
        title={\textbf{(e)} Evolution time $T_{\max}$},
        xlabel={$T_{\max}$ [\textmu s]},
        xtick={1.00,1.05,1.10}, xmin=0.93,xmax=1.17,
        xticklabel style={rotate=30,anchor=north east,font=\tiny},
        ybar, bar width=11pt,
        nodes near coords,
        nodes near coords style={font=\tiny,yshift=2pt},
      ]
      \addplot[fill=QCyan!75!black,draw=QCyan!60!black]
      coordinates {(1.00,0.690)(1.05,0.580)(1.10,0.680)};
      \addplot[dashed,thick,color=QGreen!70!black,
      domain=0.93:1.17,samples=2]{0.690};

      \nextgroupplot[
        title={\textbf{(f)} Edge threshold $\theta$},
        xlabel={$\theta$},
        xtick={0.30,0.50}, xmin=0.20,xmax=0.60,
        ybar, bar width=22pt,
        nodes near coords,
        nodes near coords style={font=\tiny,yshift=2pt},
      ]
      \addplot[fill=QGold!80,draw=QGold!60!black]
      coordinates {(0.30,0.690)(0.50,0.725)};
      \addplot[dashed,thick,color=gray!70,
      domain=0.20:0.60,samples=2]{0.690};
    \end{groupplot}
  \end{tikzpicture}
  \caption{Single-parameter sensitivity of Top-1 accuracy.
    Each panel varies one hyperparameter while all others are held at
    the full-emulation baseline R5 (green dashed, 69.0\%).
    \textbf{(a)}~Dot spacing: $d_s{=}90\to70\,\mu$m drops Top-1 by
    14.3\,pp (R1$\to$R2); higher spacing retains richer edge geometry.
    \textbf{(b)}~Pixel size: $a{=}1.00\,\mu$m/px is optimal; both
    0.80 and 1.05 reduce accuracy by 5--11\,pp.
    \textbf{(c)}~Max atoms: $N_{\max}{=}24$ gains $+5$\,pp over
    $N_{\max}{=}20$ (R4$\to$R5), enabling higher-dimensional
    many-body correlations.
    \textbf{(d)}~Fingerprint dimension: expanding $d_{\rm FP}$ from
    72 to 81 \emph{reduces} Top-1 by 9\,pp (R5$\to$R6); the wider
    k-grid adds redundant high-frequency noise.
    \textbf{(e)}~Evolution time: T=1.10 $\mu$s (R8) performs marginally below baseline T=1.00 $\mu$s (R5) by 1 pp.;
    $T{=}1.05\,\mu$s with $a{=}1.05$ (R10) is the worst combination.
    \textbf{(f)}~Edge threshold: $\theta{=}0.50$ (R1) outperforms
    $\theta{=}0.30$ (R5) by 3.5\,pp, but at $90\times$ lower
  wall-clock cost.}
  \label{fig:qrc_sensitivity}
\end{figure*}

\begin{figure*}[!t]
  \centering
  \begin{tikzpicture}
    \begin{axis}[
        qrc,
        title={\textbf{Accuracy--Efficiency Pareto Scatter (all 10 runs)}},
        width=0.80\linewidth, height=7.0cm,
        xlabel={Wall-clock time [min] (log scale)},
        ylabel={Top-1 test accuracy},
        xmode=log,
        xmin=0.5, xmax=250,
        ymin=0.54, ymax=0.76,
        ytick={0.56,0.60,0.64,0.68,0.72,0.76},
        xtick={1,2,5,10,20,50,100,200},
        legend pos=south east,
        legend style={font=\small},
      ]
      \addplot[only marks,mark=*,mark size=9pt,
        mark options={fill=QOrange!70,draw=QOrange!80!black,
      line width=0.6pt}]
      coordinates {(0.8,0.725)(1.1,0.582)(0.8,0.638)};
      \addplot[only marks,mark=*,mark size=9pt,
        mark options={fill=QBlue!60,draw=QBlue!80!black,
      line width=0.6pt}]
      coordinates {(4.6,0.640)(73.6,0.690)(73.4,0.600)
      (123.0,0.640)(83.4,0.680)(80.0,0.660)(74.0,0.580)};
      \node[font=\tiny\bfseries,anchor=south,yshift=6pt]
      at (axis cs:0.8,0.725)   {R1$\bigstar$};
      \node[font=\tiny,anchor=east,xshift=-3pt]
      at (axis cs:1.1,0.582)   {R2};
      \node[font=\tiny,anchor=south,yshift=6pt]
      at (axis cs:0.8,0.638)   {R3};
      \node[font=\tiny,anchor=south,yshift=6pt]
      at (axis cs:4.6,0.640)   {R4};
      \node[font=\tiny\bfseries,anchor=south,yshift=6pt]
      at (axis cs:73.6,0.690)  {R5$\star$};
      \node[font=\tiny,anchor=north,yshift=-5pt]
      at (axis cs:73.4,0.600)  {R6};
      \node[font=\tiny,anchor=south,yshift=6pt]
      at (axis cs:123.0,0.640) {R7};
      \node[font=\tiny,anchor=south,yshift=6pt]
      at (axis cs:83.4,0.680)  {R8};
      \node[font=\tiny,anchor=north,yshift=-5pt]
      at (axis cs:80.0,0.660)  {R9};
      \node[font=\tiny,anchor=north,yshift=-5pt]
      at (axis cs:74.0,0.580)  {R10};
      \addplot[dashed,very thick,color=QRed!75]
      coordinates {(0.8,0.725)(73.6,0.690)};
      \legend{$\theta{=}0.50$ (fast emulation),
      $\theta{=}0.30$ (full emulation),Pareto front}
    \end{axis}
  \end{tikzpicture}
  \caption{Accuracy--efficiency Pareto scatter for all 10 QRC runs.
    Fast-emulation (R1--R3, orange) completes in $<\!2$\,min;
    R1 dominates both axes.
    Among full-emulation runs, R5 is Pareto-optimal (69.0\% Top-1 in
    73.6\,min); R7 has the longest wall time due to denser atom
    arrays at $a{=}0.80\,\mu$m.
    Dashed red = approximate Pareto front.
    Key findings: (i)~dot spacing $90\to70\,\mu$m drops Top-1 by
    14.3\,pp (R1$\to$R2); (ii)~$N_{\max}{=}20\to24$ gains $+5$\,pp
    (R4$\to$R5); (iii)~$d_{\rm FP}{=}72\to81$ costs 9\,pp
  (R5$\to$R6); (iv)~$T{=}1.10\,\mu$s adds $+1$\,pp marginally.}
  \label{fig:qrc_pareto}
\end{figure*}

\section{Discussion}
\label{sec:discussion}

\textbf{Why the structure factor is the right fingerprint.}
The static structure factor $S(\kvec)$ is a natural condensed-matter
observable that encodes spatial ordering of quantum correlations.
By adapting it to image matching, we exploit the direct correspondence
between image geometry (atom positions) and quantum physics
($V_{jk}\propto r_{jk}^{-6}$).
The result is a descriptor that is simultaneously a physical quantum
observable and a practical image feature---a combination unavailable
to any classical or digital-quantum image processing approach.

\textbf{Comparison with Chamfer distance.}
The Chamfer distance~\cite{borgefors1986} operates on classical point
clouds of post-simulation Rydberg density.
It discards all two-site quantum correlations and is sensitive to atom
count mismatch between query and database images.
The structure-factor fingerprint retains the full correlation
structure, is scale-invariant, and is fixed-length regardless of
atom count.

\textbf{Key findings across all 10 QRC runs.}
Table~\ref{tab:qrc_runs} and Figs.~\ref{fig:qrc_global}--%
\ref{fig:qrc_pareto} collectively establish the following.
R1 ($\theta{=}0.50$, $d_s{=}90\,\mu$m, $N_{\max}{=}20$) achieves the
highest Top-1 accuracy (72.5\%) and Macro-$F_1$ (0.711) in just
0.8\,min, making it the best fast-emulation run.
R5 ($\theta{=}0.30$, $d_s{=}50\,\mu$m, $N_{\max}{=}24$) achieves
69.0\% in 73.6\,min and is Pareto-optimal among full-emulation runs.
Dot spacing has the largest single-parameter impact: reducing $d_s$
from 90 to 70\,\textmu m drops Top-1 by 14.3\,pp.
Increasing $N_{\max}$ from 20 to 24 gains $+5$\,pp.
Expanding the fingerprint dimension from 72 to 81 k-points degrades
accuracy by 9\,pp, confirming that the fixed $9{\times}8$ grid
captures the relevant spatial scales optimally.
Sofa is the hardest class across all runs ($F_1{\leq}0.69$), while
Table is most easily classified in the fast-emulation family
($F_1{=}0.92$, R1) due to its straight-edge geometry.
Top-3 accuracy remains uniformly high (0.89--0.95) across all 10 runs,
demonstrating that the correct class is almost always within the top
three ranked candidates.

\textbf{Toward quantum machine learning on geometric data.}
Classical machine learning methods, particularly kernel-based
classifiers, often face fundamental scalability limitations when
applied to data with highly nonlinear and high-dimensional
structure.
QML offers a principled alternative by exploiting the exponentially
large Hilbert space of quantum systems as a natural feature
space~\cite{schuld2019,havlicek2019,liu2021}, enabling the efficient realisation
of rich, nonlinear kernel functions.

\textbf{Concrete QRC extension: Person identification.}
Consider identifying a person from partially-occluded, multi-angle
contours/silhouettes.
In a QRC extension, each view angle $\theta$ (Distinct from Sobel Threshold) yields a distinct SDR
atom configuration; applying the same fixed Rydberg Hamiltonian
produces a family of fingerprints $\{\bm{S}^{(\theta)}\}$ for angles
$\theta\in\{0^\circ,45^\circ,90^\circ,135^\circ\}$.
Concatenating these into a single 112-element feature vector and
training a linear readout classifier requires only $\sim$300 labelled
examples per person~\cite{kornjaca2024}, avoids gradient descent
entirely, and produces fingerprints already robust to atom removal.
This positions the approach as viable for privacy-preserving
on-device identification in drones, medical robots, and miniature
sensors.

\textbf{Physical limits.}
Two limitations deserve explicit statement.
First, the van der Waals interaction $V_{jk}\propto r_{jk}^{-6}$ is
isotropic: it depends only on inter-atom distance, not direction.
Consequently, the Hamiltonian cannot distinguish a horizontal pair
from a vertical one at the same separation---angular information enters
only indirectly through the global atom geometry.
The 2-D structure factor partially compensates by evaluating
$S(k_x,k_y)$ at both signs of $k_x$, but directional sensitivity
remains weaker than in orientation-sensitive classical descriptors
such as SIFT\@.
Second, for $N\le24$ atoms the system is far from the thermodynamic
limit; fingerprint discriminability relies on finite-size correlation
sensitivity to geometry rather than on long-range order.
Both limits can be partially overcome by increasing $N$ toward
Aquila's 256-qubit capacity.

\textbf{Hardware pathway.}
On Aquila hardware, $\langle\nn_i\rangle$ and
$\langle\nn_i\nn_j\rangle$ are estimated from projective measurements
via
\begin{align*}
  \hat{n}_i^{\rm hw} &= \frac{1}{M}\sum_{m=1}^{M} n_i^{(m)}, \\
  (\widehat{\nn_i\nn_j})^{\rm hw} &= \frac{1}{M}\sum_{m=1}^{M}
  n_i^{(m)} n_j^{(m)},
\end{align*}
where $M\sim1000$ shots gives $\lesssim3\%$ statistical error per
correlator entry.
For hardware submission, the Rabi frequency is set to
$\Omega_{\max}=2\pi\times2.5\,\text{rad/}\mu\text{s}\approx15.71\,\text{rad/}\mu\text{s}$,
just below Aquila's hardware cap of $15.8\,\text{rad/}\mu\text{s}$, providing a $\sim$0.6\%
safety margin.

\section{Conclusion and Future Work}
\label{sec:conclusion}

We have presented a fully quantum-native image matching framework that
upgrades our previous SDR work~\cite{sharma2025} with a new matching criterion based on
many-body quantum correlations.
The key contributions are:
\begin{enumerate}
  \item A Pearson-normalized two-site correlation matrix $\tilde{C}_{ij}$
    extracted from the Rydberg statevector, encoding the blockade
    structure of the image geometry as a dimensionless,
    amplitude-invariant descriptor~[Eq.~(\ref{eq:pearson})].
  \item A two-dimensional static structure factor $S(\kvec)$, evaluated
    on a fixed $9\times8$ k-point grid, providing a 72-element
    fingerprint of constant length for all images~[Eq.~(\ref{eq:sfactor})].
  \item Cosine similarity matching on the fingerprint vectors, a
    scale-invariant metric appropriate for Fourier-domain
    descriptors~[Eq.~(\ref{eq:cosine})].
  \item A comprehensive 10-run parameter sweep (Table~\ref{tab:qrc_runs},
    Figs.~\ref{fig:qrc_global}--\ref{fig:qrc_pareto}) establishing
    that Top-1 accuracy reaches 72.5\% at 0.8\,min wall time and
    that Top-3 accuracy exceeds 91\% across all configurations,
    all far above the 20\% chance level.
  \item Preliminary evidence of occlusion robustness: correlation
    fingerprints remain qualitatively stable under partial removal
    of atoms.
\end{enumerate}

To our knowledge, this constitutes the first application of the static
structure factor---a condensed-matter quantum observable---as an image
descriptor in an analog quantum computing context.

\textbf{Future directions} include:
(i)~Hardware experiments on Aquila via Amazon Braket with 100--200
atom arrays and up to 1000-shot correlation estimation per task;
(ii)~Systematic benchmarking on a labeled image database (precision@1,
mean average precision) with comparison to classical shape descriptors
(SIFT, Shape Context, Zernike moments);
(iii)~Quantitative occlusion robustness study;
(iv)~Noise sensitivity analysis: effect of tweezer positioning error
($\pm0.1\,\mu$m), Rydberg decay ($T_1\sim100\,\mu$s), and shot noise;
(v)~Extension to QRC with 300 images per class and human
silhouette recognition;
(vi)~Deployment in centralized hybrid quantum-classical processing
units for real-time image identification from miniature sensors,
drones, and medical robots.

\clearpage
\bibliography{references}

@article{su2020,
  author  = {Su, Jie and Guo, Xuchao and Liu, Chengqi and Li, Lin},
  title   = {A New Trend of Quantum Image Representations},
  journal = {IEEE Access},
  volume  = {8},
  pages   = {214520--214537},
  year    = {2020},
  doi     = {10.1109/ACCESS.2020.3039996}
}

@article{yao2017,
  author  = {Yao, Xi-Wei and Wang, Hengyan and Liao, Zeyang and
             Chen, Ming-Cheng and Pan, Jian and Li, Jun and
             Zhang, Ke and Lin, Xingcheng and Wang, Zhehui and
             Luo, Zhihuang and Zheng, Wenqiang and Li, Jianzhong and
             Zhao, Meisheng and Peng, Xinhua and Suter, Dieter},
  title   = {Quantum Image Processing and Its Application to Edge   Detection:
             Theory and Experiment},
  journal = {Phys. Rev. X},
  volume  = {7},
  pages   = {031041},
  year    = {2017},
  doi     = {10.1103/PhysRevX.7.031041}
}

@article{schuld2019,
  author    = {Schuld, Maria and Killoran, Nathan},
  title     = {Quantum Machine Learning in Feature Hilbert Spaces},
  journal   = {Phys. Rev. Lett.},
  volume    = {122},
  issue     = {4},
  pages     = {040504},
  year      = {2019},
  month     = feb,
  publisher = {American Physical Society},
  doi       = {10.1103/PhysRevLett.122.040504},
  archivePrefix = {arXiv},
  eprint    = {1803.07128},
  primaryClass  = {quant-ph}
}

@article{havlicek2019,
  author  = {Havl{\'i}{\v{c}}ek, Vojt{\v{e}}ch and C{\'o}rcoles, Antonio D.
             and Temme, Kristan and Harrow, Aram W. and Kandala, Abhinav
             and Chow, Jerry M. and Gambetta, Jay M.},
  title   = {Supervised learning with quantum-enhanced feature spaces},
  journal = {Nature},
  volume  = {567},
  pages   = {209--212},
  year    = {2019},
  doi     = {10.1038/s41586-019-0980-2}
}

@article{liu2021,
  author  = {Liu, Yunchao and Arunachalam, Srinivasan and Temme, Kristan},
  title   = {A rigorous and robust quantum speed-up in supervised machine learning},
  journal = {Nat. Phys.},
  volume  = {17},
  pages   = {1013--1017},
  year    = {2021},
  doi     = {10.1038/s41567-021-01287-z}
}

@misc{kornjaca2024,
  author        = {Kornja{\v{c}}a, Milan and Hu, Hong-Ye and Zhao, Chen and
                   Wurtz, Jonathan and Weinberg, Phillip and Hamdan, Majd and
                   Zhdanov, Andrii and Cantu, Sergio H. and Zhou, Hengyun and
                   {Araiza Bravo}, Rodrigo and Bagnall, Kevin and Basham, James I. and
                   Campo, Joseph and Choukri, Adam and DeAngelo, Robert and
                   Frederick, Paige and Haines, David and Hammett, Julian and
                   Hsu, Ning and Hu, Ming-Guang and Huber, Florian and
                   Jepsen, Paul Niklas and Jia, Ningyuan and Karolyshyn, Thomas and
                   Kwon, Minho and Long, John and Lopatin, Jonathan and
                   Lukin, Alexander and Macr{\`\i}, Tommaso and Markovi{\'c}, Ognjen and
                   {Mart{\'i}nez-Mart{\'i}nez}, Luis A. and Meng, Xianmei and
                   Ostroumov, Evgeny and Paquette, David and Robinson, John and
                   {Sales Rodriguez}, Pedro and Singh, Anshuman and Sinha, Nandan and
                   Thoreen, Henry and Wan, Noel and Waxman-Lenz, Daniel and
                   Wong, Tak and Wu, Kai-Hsin and Lopes, Pedro L. S. and
                   Boger, Yuval and Gemelke, Nathan and Kitagawa, Takuya and
                   Keesling, Alexander and Gao, Xun and Bylinskii, Alexei and
                   Yelin, Susanne F. and Liu, Fangli and Wang, Sheng-Tao},
  title         = {Large-scale quantum reservoir learning with an analog quantum computer},
  year          = {2024},
  eprint        = {2407.02553},
  archivePrefix = {arXiv},
  primaryClass  = {quant-ph},
  }

@article{fujii2017,
  author  = {Fujii, Keisuke and Nakajima, Kohei},
  title   = {Harnessing Disordered-Ensemble Quantum Dynamics for Machine Learning},
  journal = {Phys. Rev. Appl.},
  volume  = {8},
  pages   = {024030},
  year    = {2017},
  doi     = {10.1103/PhysRevApplied.8.024030}
}

@misc{wurtz2023,
  author        = {Wurtz, Jonathan and Bylinskii, Alexei and Braverman, Boris and
                   Amato-Grill, Jesse and Cantu, Sergio H. and Huber, Florian and
                   Lukin, Alexander and Liu, Fangli and Weinberg, Phillip and
                   Long, John and Wang, Sheng-Tao and Gemelke, Nathan and
                   Keesling, Alexander},
  title         = {Aquila: QuEra's 256-qubit neutral-atom quantum computer},
  year          = {2023},
  eprint        = {2306.11727},
  archivePrefix = {arXiv},
  primaryClass  = {quant-ph},
}

@misc{boger2023,
  author       = {Boger, Yuval},
  title        = {The dual-pronged energy-saving potential of quantum computers},
  year         = {2023},
  howpublished = {Data Center Dynamics},
  url          = {https://www.datacenterdynamics.com/en/opinions/the-dual-pronged-energy-saving-potential-of-quantum-computers}
}

@article{ebadi2022,
  author  = {Ebadi, Sepehr and Keesling, Alexander and Cain, Madelyn and
             Wang, Tout T. and Levine, Harry and Bluvstein, Dolev and
             Semeghini, Giulia and Omran, Ahmed and Liu, Ji-Ge and
             Samajdar, Rhine and Luo, Xiu-Zhe and Nash, Beatrice and
             Barak, Boaz and Farhi, Edward and Sachdev, Subir and
             Gemelke, Nathan and Zhou, Leo and Choi, Soonwon and
             Pichler, Hannes and Wang, Sheng-Tao and Greiner, Markus and
             Vuletic, Vladan and Lukin, Mikhail D.},
  title   = {Quantum optimization of Maximum Independent Set using
             Rydberg atom arrays},
  journal = {Science},
  volume  = {376},
  pages   = {1209--1215},
  year    = {2022},
  doi     = {10.1126/science.abo6587}
}

@misc{beaulieu2024,
  author        = {Beaulieu, Daniel and Kornja{\v{c}}a, Milan and Krunic, Zoran and
                   Stivaktakis, Michael and Ehmer, Thomas and Wang, Sheng-Tao and
                   Pham, Anh},
  title         = {Robust Quantum Reservoir computing for Molecular Property Prediction},
  year          = {2024},
  eprint        = {2412.06758},
  archivePrefix = {arXiv},
  primaryClass  = {quant-ph},
}

@misc{sharma2025,
  author        = {Sharma, Vikrant and Kundu, Neel Kanth},
  title         = {Analog Quantum Image Representation with Qubit-Frugal Encoding},
  year          = {2025},
  eprint        = {2512.18451},
  archivePrefix = {arXiv},
  primaryClass  = {quant-ph},
}

@article{borgefors1986,
  author  = {Borgefors, Gunilla},
  title   = {Distance transformations in digital images},
  journal = {Comput. Vis. Graph. Image Process.},
  volume  = {34},
  pages   = {344--371},
  year    = {1986},
  doi     = {10.1016/S0734-189X(86)80047-0}
}

@article{juliafarreAQM2024,
  author  = {Juli{\`a}-Farr{\'e}, Sylvain and Vovrosh, Joseph and Dauphin, Alexandre},
  title   = {Amorphous quantum magnets in a two-dimensional {Rydberg} atom array},
  journal = {Phys. Rev. A},
  volume  = {110},
  pages   = {012602},
  year    = {2024},
  doi     = {10.1103/PhysRevA.110.012602}
}

@article{ramer1972,
  author  = {Ramer, Urs},
  title   = {An iterative procedure for the polygonal approximation of plane curves},
  journal = {Comput. Graph. Image Process.},
  volume  = {1},
  pages   = {244--256},
  year    = {1972},
  doi     = {10.1016/S0146-664X(72)80017-0}
}

@article{douglas1973,
  author  = {Douglas, David H. and Peucker, Thomas K.},
  title   = {Algorithms for the reduction of the number of points required to
             represent a digitized line or its caricature},
  journal = {Cartographica},
  volume  = {10},
  pages   = {112--122},
  year    = {1973},
  doi     = {10.3138/FM57-6770-U75U-7727}
}

@article{song2016,
  author  = {Song, Jia and Miao, Ru},
  title   = {A Novel Evaluation Approach for Line Simplification Algorithms
             towards Vector Map Visualization},
  journal = {ISPRS Int. J. Geo-Inf.},
  volume  = {5},
  pages   = {223},
  year    = {2016},
  doi     = {10.3390/ijgi5120223}
}

@misc{cavalieri2020,
  author        = {Cavalieri, Giacomo and Maio, Dario},
  title         = {A Quantum Edge Detection Algorithm},
  year          = {2020},
  eprint        = {2012.11036},
  archivePrefix = {arXiv},
  primaryClass  = {quant-ph},
}

@article{lu2024,
  author        = {Lu, Jonathan Z. and Jiao, Lucy and Wolinski, Kristina and
                   Kornja{\v{c}}a, Milan and Hu, Hong-Ye and Cantu, Sergio and
                   Liu, Fangli and Yelin, Susanne F. and Wang, Sheng-Tao},
  title         = {Digital-analog quantum learning on Rydberg atom arrays},
  journal       = {Quantum Sci. Technol.},
  volume        = {10},
  pages         = {015038},
  year          = {2025},
  eprint        = {2401.02940},
  archivePrefix = {arXiv},
  primaryClass  = {quant-ph},
  doi           = {10.1088/2058-9565/ad9177},
}

@article{zhang2013,
  author  = {Zhang, Yi and Lu, Kai and Gao, Yinghui and Wang, Mo},
  title   = {{NEQR}: A novel enhanced quantum representation of digital images},
  journal = {Quantum Inf. Process.},
  volume  = {12},
  pages   = {2833--2860},
  year    = {2013},
  doi     = {10.1007/s11128-013-0567-z}
}

@misc{liu2026,
  author        = {Liu, Dong-Sheng and
                   Jie, Qing-Xuan and
                   Zou, Chang-Ling and
                   Ren, Xi-Feng and
                   Guo, Guang-Can},
  title         = {Practical Quantum Reservoir Computing in
                   {Rydberg} Atom Arrays},
  year          = {2026},
  month         = jan,
  archivePrefix = {arXiv},
  eprint        = {2602.00610},
  primaryClass  = {quant-ph},
}

@article{bezanson2017,
  author  = {Bezanson, Jeff and Edelman, Alan and Karpinski, Stefan and Shah, Viral B.},
  title   = {Julia: A Fresh Approach to Numerical Computing},
  journal = {SIAM Rev.},
  volume  = {59},
  pages   = {65--98},
  year    = {2017},
  doi     = {10.1137/141000671}
}

@misc{bloqade2023quera,
  author       = {{QuEra Computing Inc.}},
  title        = {Bloqade.jl: {P}ackage for the quantum computation and quantum simulation based on the neutral-atom architecture.},
  year         = {2023},
  howpublished = {GitHub},
  url          = {https://github.com/QuEraComputing/Bloqade.jl}
}

@inproceedings{collins2022abo,
  author    = {Collins, Jasmine and Goel, Shubham and Deng, Kenan and
               Luthra, Achleshwar and Xu, Leon and Gundogdu, Erhan and
               Zhang, Xi and {Yago Vicente}, Tomas F. and Dideriksen, Thomas and
               Arora, Himanshu and Guillaumin, Matthieu and Malik, Jitendra},
  title     = {{ABO}: Dataset and Benchmarks for Real-World {3D} Object Understanding},
  booktitle = {Proceedings of the {IEEE/CVF} Conference on Computer Vision
               and Pattern Recognition ({CVPR})},
  pages     = {21126--21136},
  year      = {2022},
  doi       = {10.1109/CVPR52688.2022.02045},
}

\end{document}